\documentclass[a4paper,fleqn,usenatbib,useAMS]{mnras}

\usepackage{array}
\usepackage{color}
\usepackage{subfigure}
\usepackage{graphicx}	
\usepackage{amsmath}	
\usepackage{amssymb}	
\usepackage{multicol}        
\usepackage{bm}		
\usepackage{pdflscape}	
\usepackage{ulem} 

\usepackage[T1]{fontenc}
\usepackage{ae,aecompl}

\usepackage{newtxtext,newtxmath}

\title[Cosmic HI density]
{An accurate low-redshift measurement of the cosmic neutral hydrogen density}

\author[Wenkai Hu et al.]
{Wenkai Hu$^{1,2,3,4}$\thanks{Contact e-mail: \href{wkhu@nao.cas.cn}{wkhu@nao.cas.cn}},
Laura Hoppmann$^{1}$, 
Lister Staveley-Smith$^{1,4}$, 
Katinka Ger$\rm\acute{e}$b$^{5}$, 
\newauthor
Tom Oosterloo$^{6,7}$,
Raffaella Morganti$^{4,6,7}$,
Barbara Catinella$^{1,4}$,
Luca Cortese$^{1,4}$,
\newauthor
Claudia del P. Lagos$^{1,4}$,
Martin Meyer$^{1,4}$
\\
$^{1}$ International Centre for Radio Astronomy Research (ICRAR), M468, University of Western Australia, 35 Stirling Hwy, WA 6009, Australia\\
$^{2}$ Key Laboratory of National Astronomical Observatories, Chinese Academy of Sciences, Beijing 100012, China\\
$^{3}$ School of Astronomy and Space Science, University of Chinese Academy of Sciences, Beijing 100049, China\\
$^{4}$ ARC Centre of Excellence for All Sky Astrophysics in 3 Dimensions (ASTRO 3D)\\
$^{5}$ Centre for Astrophysics and Supercomputing, Swinburne University of Technology, Hawthorn, VIC 3122, Australia\\
$^{6}$ ASTRON, the Netherlands Institute for Radio Astronomy, Postbus 2, 7990 AA, Dwingeloo, The Netherlands\\
$^{7}$ Kapteyn Astronomical Institute, University of Groningen, P.O. Box 800, 9700 AV Groningen, The Netherlands\\
}

\date{Last updated XXX; in original form XXX}

\pubyear{2018}

\begin{document}
\label{firstpage}
\pagerange{\pageref{firstpage}--\pageref{lastpage}}
\maketitle

\begin{abstract}
Using a spectral stacking technique, we measure the neutral hydrogen (HI) properties of a sample of galaxies at $z < 0.11$ across 35 pointings of the Westerbork Synthesis Radio Telescope (WSRT). The radio data contains 1,895 galaxies with redshifts and positions known from the Sloan Digital Sky Survey (SDSS). We carefully quantified the effects of sample bias, aperture used to extract spectra, sidelobes and weighting technique and use our data to provide a new estimate for the cosmic HI mass density. We find a cosmic HI mass density of $\Omega_{\rm HI} = (4.02 \pm 0.26)\times 10^{-4} h_{70}^{-1}$ at $\langle z\rangle = 0.066$, consistent with measurements from blind HI surveys and other HI stacking experiments at low redshifts. The combination of the small interferometer beam size and the large survey volume makes our result highly robust against systematic effects due to confusion at small scales and cosmic variance at large scales. Splitting into three sub-samples with $\langle z\rangle$ = 0.038, 0.067 and 0.093 shows no significant evolution of the HI gas content at low redshift. 
\end{abstract}

\begin{keywords}
galaxies: evolution - galaxies: ISM - radio lines: galaxies
\end{keywords}




\section{Introduction}
To fully understand the formation and evolution of galaxies, it is important to study the accretion of gas from the intergalactic medium (IGM), galaxy mergers and galaxy interaction, and the depletion of gas through galactic fountains and outflow processes \citep{2005MNRAS.363....2K,2008A&ARv..15..189S,2010MNRAS.404.1464M}. Cool gas drives star formation in galaxies as shown by the correlation between star-formation surface density ($\Sigma_{\rm SFR}$) and HI surface density ($\Sigma_{\rm HI}$) \citep{1959ApJ...129..243S,1998ApJ...498..541K}, and the even tighter correlation with molecular hydrogen surface density ($\Sigma_{\rm H_{2}}$) \citep{2008AJ....136.2846B,2010ApJ...722.1699S}. Whilst the latter provides evidence for the important role of molecular clouds in controlling star formation \citep{2005ARA&A..43..677S}, because of the relatively short gas consumption time scales, it is the large-scale net inflow and condensation of cool gas which eventually forms the massive molecular clouds prior to star formation. Therefore study of both the atomic and molecular phases of cool gas in galaxies is crucial for the understanding of their star formation history.

There are a number of observation techniques we can use to measure HI gas content.  At high redshifts, the damped Lyman-$\alpha$ (DLA) systems seem to indicate large reservoirs of HI whose column density can be deduced from DLA absorption profiles, thereby allowing determination of the cosmic HI mass density. At $z > 1.65$, many DLA surveys have therefore been used to measure the cosmic HI gas density \citep{1991ApJS...77....1L,2005ApJ...635..123P,2009A&A...505.1087N,2012A&A...547L...1N,2010ApJ...721.1448S,2013A&A...556A.141Z,2015MNRAS.452..217C,2016ApJ...818..113N,2017MNRAS.466.2111B}. Their results show a significant evolution of HI gas content over cosmic time and that there is more HI gas at high redshifts. At $z < 1.65$, Lyman-$\alpha$ absorption is only detected at ultraviolet (UV) wavelengths, so can only be observed with space-based telescopes. \citet{2006ApJ...636..610R,2017MNRAS.471.3428R} have identified candidate DLA systems through their metal absorption lines in the redshift range $0.11 < z < 1.65$. Their results indicate no clear evolution of cosmic HI gas density. However, the low incidence of DLAs per unit redshift at intermediate redshifts give rise to significant statistical uncertainties.

In the local Universe, the HI content is conveniently measured through the direct detection of the 21-cm hyperfine emission line. The large instantaneous field of view provided by modern multibeam receivers has made blind, large-area HI surveys possible. The HI Parkes All-Sky Survey (HIPASS) \citep{2001MNRAS.322..486B} has detected HI emission from 5317 galaxies at $0 < z < 0.04$ over a sky area of 21,341 deg$^{2}$ \citep{2004MNRAS.350.1195M,2006MNRAS.371.1855W}, and the Arecibo Legacy Fast ALFA (ALFALFA) survey \citep{2005AJ....130.2598G} has detected $\sim$ 31500 galaxies out to $z = 0.06$ over a sky area of approximately 7000 deg$^{2}$ \citep{2018ApJ...861...49H}. These large-area surveys allow for accurate measurement of the local HI mass function and the cosmic HI gas density. The measurements of HI density from these surveys are reasonably consistent with each other \citep{2005MNRAS.359L..30Z,2010ApJ...723.1359M,2018MNRAS.tmp..502J}. However, directly measuring 21-cm emission of more distant individual galaxies is difficult with the current generation of single-dish radio telescopes, so this approach is limited to low redshift.

Individual deep 21-cm pointings have proven the feasibility of detecting HI galaxies outside the local Universe and up to $z \approx 0.3$ \citep{2008ApJ...685L..13C,2001Sci...293.1800Z,1538-4357-668-1-L9,2016ApJ...824L...1F}. However, in order to increase the chance of detection, the observed areas are often pre-selected. For example, \citet{2015MNRAS.446.3526C} detected 39 galaxies up to $z = 0.25$ with the 305-m Arecibo telescope, selecting them by presence of H$\alpha$ emission, disk morphology and isolation. \citet{2001Sci...293.1800Z} and \citet{1538-4357-668-1-L9} targeted galaxies in clusters at $z \approx 0.2$ with the Westerbork Synthesis Radio Telescope (WSRT). These samples are biased towards bright galaxies with high optical surface brightness, or in dense regions.

However, blind surveys to higher redshifts are time consuming. For example, the Arecibo Ultra Deep Survey (AUDS) \citep{2011ApJ...727...40F,2015MNRAS.452.3726H}  has so-far detected 103 galaxies with 400 hrs of integration time in the redshift range of $0 < z < 0.16$. The Cosmological Evolution Survey (COSMOS) HI Large Extragalactic Survey (CHILES) over the redshift range $z = 0$ - 0.45 \citep{2041-8205-770-2-L29,2016ApJ...824L...1F} will be able to detect up to 300 galaxies with 1000 hours of observation time on the Very Large Array (VLA). However, even with such large integration times, these surveys have been limited to very small sky areas (1.35 deg$^{2}$ for AUDS and 0.3 deg$^{2}$ for CHILES), resulting in small effective volumes and large cosmic variance. 

Next generation telescopes SKA pathfinder such as Australian Square Kilometre Array Pathfinder (ASKAP) \citep{2008ExA....22..151J,2009pra..confE..15M}, MeerKAT \citep{2012IAUS..284..496H}, Five-hundred-meter Aperture Spherical radio Telescope (FAST) \citep{2011IJMPD..20..989N,2008MNRAS.383..150D} and WSRT/Aperture Tile in Focus (APERTIF) \citep{2009wska.confE..70O}
will enable large-area surveys to significant depths. But less direct methods for measuring HI gas content at higher redshifts are also available using the technique of spectral stacking \citep{2001A&A...372..768C}. The technique combines a large number of rest-frame spectra extracted from the radio data with redshifts and positions from optical catalogues. This allows the noise to be averaged down, and recovers a more significant spectral line signal, but averaged over a large sample of galaxies. By potentially accessing a larger number of galaxies, HI stacking can provide significantly large volumes, and much smaller cosmic variance.

Studies using the spectral stacking technique for galaxies outside the local Universe include those of \citet{1538-4357-668-1-L9} and \citet{2009MNRAS.399.1447L} who examined galaxies in cluster environments out to $z = 0.37$. Other observations have been used to study the properties of nearby galaxies, for example the relation between the HI content of a galaxy and its bulge \citep{2011MNRAS.411..993F} and correlations between the HI content, stellar mass and environment \citep{2012MNRAS.427.2841F,2015MNRAS.452.2479B,2018MNRAS.473.1868B}, as well as the influence of an active galactic nucleus (AGN) \citep{2011arXiv1104.0414F,2013A&A...558A..54G}. The first attempt to use stacking to calculate the cosmic HI gas density $\rm \Omega_{HI}$, was presented by \citet{2007MNRAS.376.1357L} in the redshift range $0.218 < z < 0.253$ using the Giant Metrewave Radio Telescope (GMRT). A more recent HI stacking experiment was carried out by \citet{2013MNRAS.433.1398D} using HIPASS data and new observations from the Parkes telescope combined them with $\sim 18,300$ redshifts from the Two-Degree Field Galaxy Redshift Survey (2dFGRS) to obtain high signal-to-noise ratio detections out to a redshift of $z = 0.13$. 

\citet{2013MNRAS.435.2693R} used data from WSRT and stacked a significantly smaller sample of 59 galaxies at $z \approx 0.1$ and 96 galaxies at $z \approx 0.2$. \citet{2016MNRAS.460.2675R} cross-matched the zCOSMOS-bright catalogue with data from GMRT, obtaining a 474 galaxy sample at $z \approx 0.37$. With the stacking technique, they made a 3$\sigma$ detection of average HI mass. \citet{2018MNRAS.473.1879R} used observations made with the GMRT to probe the HI gas content of 165 field galaxies in the VIMOS VLT Deep Survey (VVDS) 14h field at $z \approx 0.32$, resulting in a measurement of HI mass with a significance of 2.8$\sigma$. \citet{2016ApJ...818L..28K} used the GMRT to stack HI emission from massive star-forming galaxies at $z\approx 1.18 –- 1.34$, the highest redshift at which stacking has been attempted.

The technique of `intensity mapping' can also be used to extend the HI survey limit to higher redshifts. Similar to stacking, this involves measuring the cross-power between radio and optical surveys \citep{doi:10.1111/j.1745-3933.2008.00581.x}, but uses the bulk emission fluctuations due to galaxy clustering over the surveyed region instead of individual galaxies. Observations conducted with the Green Bank Telescope \citep{2010arXiv1007.3709C,2013ApJ...763L..20M}, spanning the redshift range $0.6 < z < 1$ have highlighted the potential power of this technique. However, the accuracy of cosmic HI density measurements remains low, and there is a dependence on simulations of the wavelength-dependent bias of galaxies at optical and radio wavelengths \citep{2017MNRAS.470.3220W}. 

In this paper, we foreshadow some of the techniques which will be utilised in the future SKA pathfinder surveys to bridge the redshift gap $0.2 < z < 1.65$. We achieve this by using an interferometer in order to reduce problems arising from confusion that affect single-dish data. But we also cover a wide field of view by using multiple pointing centres in order to reduce cosmic variance, which has otherwise affected deep interferometer surveys. We obtain the radio data from WSRT \citep{2015A&A...580A..43G} and use a corresponding optical catalog from SDSS \citep{2000AJ....120.1579Y} containing 1895 galaxies within the sampled redshift range. Sample selection is not biased by environment, star formation, or any particular physical characteristic other than the optical magnitude limits of the SDSS. 

Section~\ref{sec:data} presents the observational data used in this paper. In Section~\ref{sec:script} we present the spectral extraction and stacking methodology. In Section~\ref{sec:results} we measure average HI mass and HI mass-to-light ratio for the sample, and various sub-samples in redshift and luminosity. In Section~\ref{sec:density} we describe our measurement of $\Omega_{HI}$ and compare with existing results in the literature.
Throughout this paper we use H$_{\circ} = 70$ km s$^{-1}$ Mpc$^{-1}$, $\Omega_{\rm m} = 0.3$ and $\Omega_{\Lambda} = 0.7$.

\section{Data}
\label{sec:data}
\begin{figure*}
    \centering
    \includegraphics[width=14.5cm]{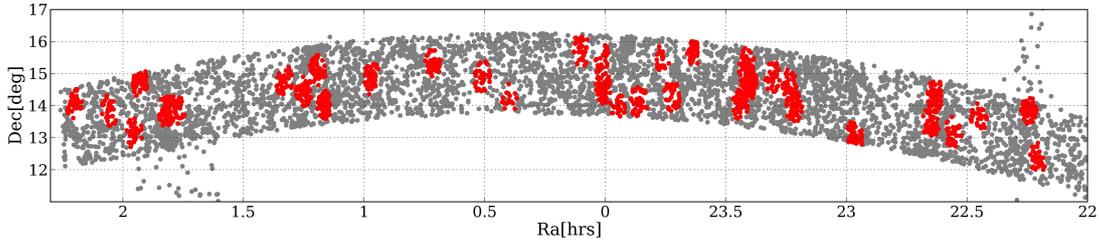}
    \caption{Positions of the galaxies contained within the 35 individual WSRT pointings observed with the WSRT (red). Other galaxies in the GALEX/SDSS strip  are shown in grey.}
    \label{pointings}
\end{figure*}

The HI observations were made using the Westerbork Synthesis Radio Telescope (WSRT). Thirty six pointing positions were selected according to the overall WSRT schedule with the only main constraint being that the sky overlap with footprint of the Galaxy Evolution eXplorer (GALEX) survey \citep{2005ApJ...619L...1M} and Sloan Digital Sky Survey (SDSS) South Galactic Cap region ($\rm 21^h < RA < 2^h$ and $\rm 10^{\circ} < Dec < 16^{\circ}$). 351 hours of observation time were used to observe the region, with each pointing observed for between 5 hr and 12 hr. Data from one of the pointings were discarded due to bad data quality. The sky region covered by the remaining 35 pointings is shown in Figure~\ref{pointings}.

\begin{figure}
    \centering
    \includegraphics[width=8cm]{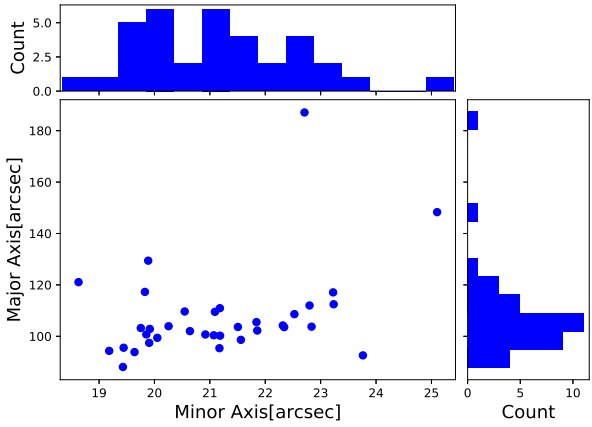}
    \includegraphics[width=8cm]{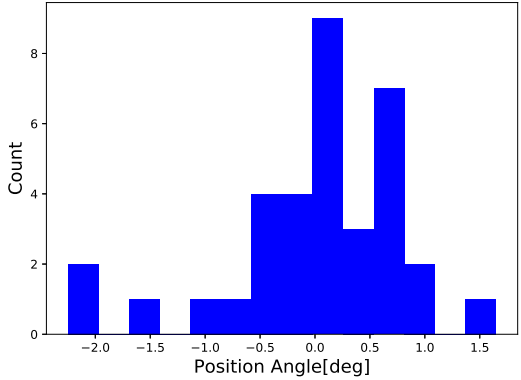}
    \caption{A histogram of the major axis, minor axis and position angle of the 35 synthesized beams, obtained from a Gaussian fit to the dirty beam point spread functions. One beam has a very large major axis ($\sim 180$ arcsec) due to poorer uv coverage.}
    \label{psf_histogram}
\end{figure}

The half-power beam width (HPBW) of the WSRT is 35 arcmin at the observing frequency, and the average synthesized beam size is $108\arcsec \times 22\arcsec$. Figure~\ref{psf_histogram} shows a histogram of major axis, minor axis and position angles of the synthesized beams for the 35 pointings. The data were reduced and self-calibrated using the radio astronomy data reduction package {\sc miriad} \citep{1995ASPC...77..433S}. The data were flagged to reduce the contamination by radio frequency interference (RFI).

The reduced data cubes have a size of 601$\times$601 pixels with a pixel size of $3\arcsec \times 3\arcsec$. The data consist of 8$\times$20 MHz bands, each with 128 channels and two polarisations. Each channel is 0.15625 MHz wide,  corresponding to $\sim 33$ km s$^{-1}$ at $z=0$ and $\sim 37$ km s$^{-1}$ at $z=0.11$. The rms was typically 0.2 mJy beam$^{-1}$ per 0.15625 MHz channel for each field, independent of frequency. Each frequency band overlaps by 3 MHz resulting in an overall frequency range of 1.406 GHz to 1.268 GHz, corresponding to a redshift range of $0.01 < z < 0.12$. However, due to stronger RFI at higher redshift we set an upper redshift limit of $z = 0.11$.

Accurate measurements of redshift and spacial positions are indispensable for stacking. We use SDSS DR9 as the optical catalogue for our stacking analysis. SDSS has a typical redshift error of $\sim 60$ km s$^{-1}$ and a spectral density of 60 - 100 deg$^{-2}$ ($z < 0.12$) in the region we selected for the HI observations. With the target selection algorithm described in \citet{2002AJ....124.1810S}, the SDSS sample has a completeness which exceeds 99$\%$ (excluding fibre collisions). The sample appears to be complete for a star formation rate above 10$^{-2}$M$_{\sun}$~yr$^{-1}$ for $z<0.06$. The luminosities used in this paper are calculated from the SDSS $r$-band magnitudes, applying $k$-corrections \citep{2012MNRAS.419.1727C}.

\begin{figure}
    \centering
    \includegraphics[width=8cm]{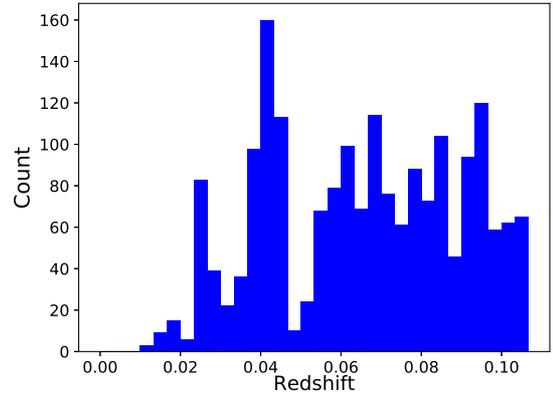}
    \caption{Redshift distribution of the SDSS spectroscopic sample contained within the 35 pointings. The width of the redshift bins is 0.003. The selected sample has an lower redshift limit of $z = 0.01$ and upper redshift limit of $z = 0.11$. The mean redshift of the sample is $\langle z\rangle = 0.066$.}
    \label{redshift_distribution}
\end{figure}

\begin{figure}
    \centering
    \includegraphics[width=8cm]{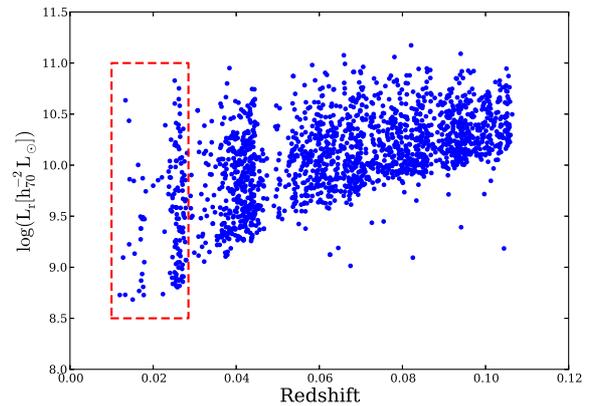}
    \caption{A plot of the $r$-band luminosity as a function of redshift $z$, for the SDSS sample. The red-dashed rectangular encloses a volume-limited sub-sample.}
    \label{redshift-L_r}
\end{figure}

By cross-matching our radio data with the SDSS catalog, we obtain a sample of 1895 galaxies spanning the redshift range $0.01 < z < 0.11$ (Figure~\ref{redshift_distribution}) and within the radius of the pointings at which the normalized primary beam response drops to 0.1. We refer to this as the magnitude-limited sample, only including galaxies with $r$-band magnitude brighter than 17.77. It has a mean redshift of $\langle z \rangle = 0.066$. To measure the HI density with a sample less biased by magnitude, we also created a volume-limited sample with $z \le 0.0285$, which has 149 galaxies in total and a mean redshift of $\langle z\rangle$ = 0.024. The volume-limited sample is complete for $r$-band luminosities  $>10^{8.68}$L$_{\odot}$. Figure~\ref{redshift-L_r} shows the $r$-band luminosity distribution as a function of redshift with the volume-limited sub-sample highlighted. 

\section{Stacking Analysis}
\label{sec:script}
\subsection{HI Mass Spectra}
\label{sec:mainprocedure}
The stacking technique used in this paper is similar to that described in \citet{2013A&A...558A..54G}. Spectra were extracted from the data cubes over an extended region around the SDSS position. After extensive tests, we find the region with aperture radius of 35kpc gives best stacking results(see Section~\ref{sec:aperture}). The spatially-integrated spectrum was calculated from:
\begin{eqnarray}
    S_{\nu} = \frac{\Sigma_{x}\Sigma_{y}S_{\nu}(x,y)}{\Sigma_{x}\Sigma_{y}B(x,y)},
    \label{spectrum}
\end{eqnarray}
where $S_{\nu}(x,y)$ is the flux density at pixel position $(x,y)$ and $B(x,y)$ is the normalized synthesized beam response (centred on the SDSS position) at the same pixel position. After this, a second-order baseline was fitted to remove residual continuum (excluding a velocity range of 500 km s$^{-1}$ around the expected spectral location of the SDSS galaxy), and the spectra were de-redshifted. The barycentric frequency is converted from the observed to the rest frame by $\nu_{\rm res} = \nu_{\rm obs}(1+z)$. As the channel width is also broadened in this process, HI flux density is conserved by applying the corresponding correction:
\begin{eqnarray}
    S_{\nu_{\rm res}} = \frac{S_{\nu_{\rm obs}}}{(1+z)} .
    \label{redshift_shift}
\end{eqnarray}
After shifting to the rest frame, the flux spectra were converted into mass spectra using the following relation:
\begin{eqnarray}
    m_{\rm HI}(\nu) = 4.98\times10^{7} S_{\nu} D_{L}^{2}f^{-1} ,
    \label{mass_spectrum}
\end{eqnarray}
where $S_{\nu}$ is the de-redshifted HI flux density in Jy, $D_{L}$ is the luminosity distance in Mpc, $f$ is the normalised primary beam response, and $m_{\rm HI}$ is in units of M$_{\sun}$ MHz$^{-1}$.

We introduce a weight factor which depends on the primary beam response $f$, the luminosity distance $D_{L}$, as well as the rms noise of the flux density spectra $\sigma$. The weight of $i$-th galaxy is expressed as:
\begin{eqnarray}
    w_{i} = f^{2}D_{L}^{-\gamma}\sigma^{-2},
    \label{weight}
\end{eqnarray}
where large values of $\gamma$ give larger weight to nearby galaxies, and small values give more weight to distant galaxies. The effect of the weight factor on the results is considered later. The averaged final stacked spectrum is obtained from:
\begin{eqnarray}
    \langle m_{\rm HI}(\nu)\rangle = \frac{\sum_{i=1}^{n}w_{i}m_{\rm HI,i}}{\sum_{i=1}^{n}w_{i}}.
    \label{weighted_spectrum}
\end{eqnarray}
The integrated HI mass of a stack, or $\langle M_{\rm HI}\rangle$, is then defined as the integral along the frequency axis over the mass spectrum:
\begin{eqnarray}
    M_{\rm HI} = \int_{\nu_{0}-\Delta\nu}^{\nu_{0}+\Delta\nu}\langle m_{\rm HI}(\nu)\rangle d\nu,
    \label{integrated_mass}
\end{eqnarray}
where $\nu_{0}$ refers to 1420.406 MHz and $\Delta\nu$ is large enough to capture all flux from the stack (we will later use $\Delta\nu = 1.5$ MHz, corresponding to $\pm 317$ km s$^{-1}$). 

We estimate the error of the HI mass measurement through jackknife resampling. From the total sample of $n$ spectra, $n/20$ randomly selected spectra are removed at a time to construct 20 jackknife samples, from which 20 mass spectra are obtained.

The jackknife estimate of the true variance of the measured value of the mass spectrum at a given frequency is then given by:
\begin{eqnarray}
     \sigma^2(\langle{m}_{\rm HI}\rangle) = \frac{19}{20}\sum_{j=1}^{20}(\langle{m}_{\rm HI}\rangle-\langle{m}_{\rm HI}^{j}\rangle)^{2},
    \label{jackknife}
\end{eqnarray}
where the $\langle{m}_{\rm HI}\rangle$ refers to the averaged HI mass spectrum from the original sample. 

We can also measure $\langle M_{\rm HI}/L\rangle$ and its error by stacking the individual $M_{\rm HI}/L$ spectra. We do this via Equation~\ref{weighted_spectrum} and \ref{integrated_mass}, with $M_{\rm HI}$ replaced by $M_{\rm HI}/L$.

\subsection{Weighting}
\label{sec:weight}
\begin{figure}
    \centering
    \includegraphics[width=8cm]{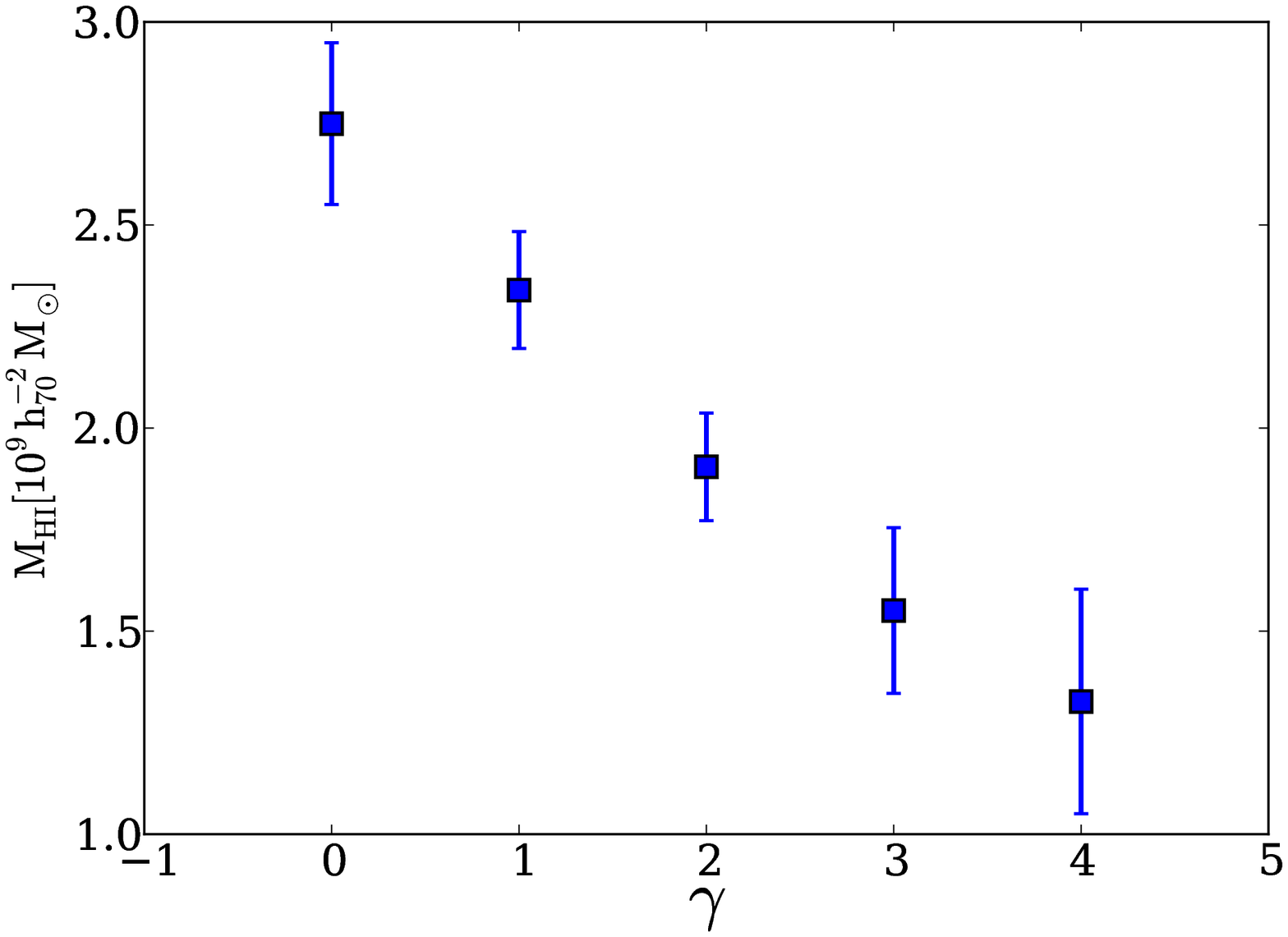}
    \includegraphics[width=8cm]{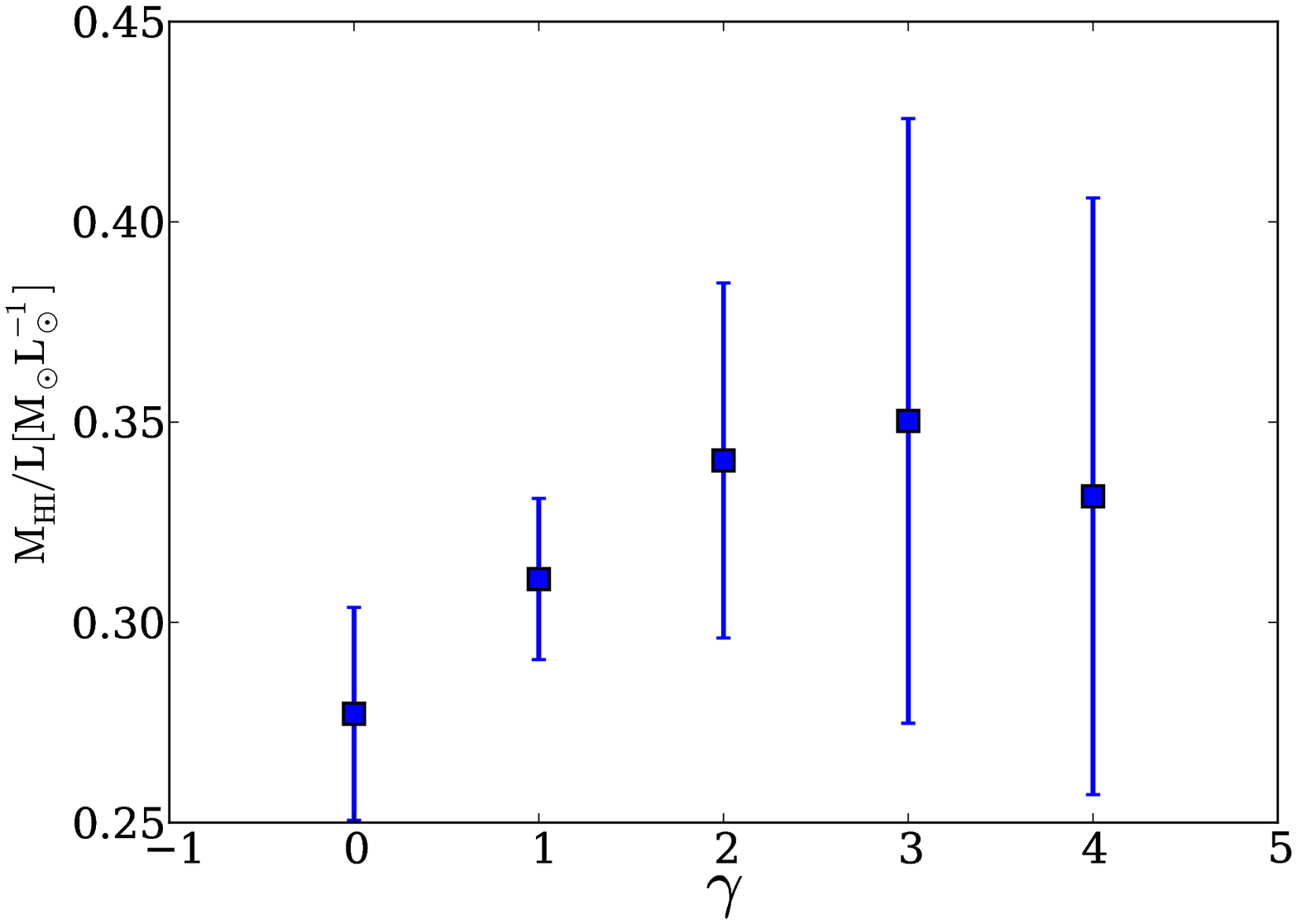}
    \caption{Average HI mass $M_{\rm HI}$ (top) and average HI mass-to-light ratio $M_{\rm HI}/L$ (bottom) as a function of weight parameter $\gamma$. The errors are obtained by jackknife sampling.}
    \label{weight_plot}
\end{figure}

In order to investigate the effect of different weights on the results, we explore the range $0 \le \gamma \le 4$. As shown in Figure~\ref{weight_plot} and Table~\ref{weight_table}, highly significant values for $\langle M_{\rm HI}\rangle$ are obtained for all weighting parameters. $\langle M_{\rm HI}\rangle$ monotonically decreases as $\gamma$ increases, reflecting the lower HI mass of nearby galaxies. Similarly, $\langle M_{\rm HI}/L_{r}\rangle$ increases with $\gamma$, although the variation is somewhat less significant. The highest overall S/N occurs when $\gamma = 1$. 
As shown in Table~\ref{weight_table}, the weighted mean redshift decreases with increasing $\gamma$. The measurements at $\gamma = 1$ are more representative of the entire sample: larger $\gamma$ gives more weight to nearby galaxies; smaller $\gamma$ gives too much weight to low S/N ratio measurements of distant galaxies.

\begin{table}
	\centering
	\caption{Mean HI mass $M_{\rm HI}$ and mean $r$-band HI mass-to-light ratio $M_{\rm HI}/L_{r}$ for the magnitude-limited sample using different values of the weight parameter $\gamma$. The weighted mean redshift is also given.}
	\label{weight_table}
	\begin{tabular}{ccccccc} 
		\hline
		$\gamma$ & <z> & <$\rm M_{HI}$> & $\rm S/N(M_{HI})$ & <$\rm M_{HI}/L_{r}$> & $\rm S/N(M_{HI}/L)$\\
        & &($10^{9}h_{70}^{-2}M_{\odot}$)& &($M_{\odot}/L_{\odot}$)& \\
		\hline
		0 &0.062 & 2.75 $\pm$ 0.20 & 13.8 & 0.28 $\pm$ 0.03 & 10.4\\
		1 & 0.051 & 2.34 $\pm$ 0.14 & 16.3 & 0.31 $\pm$ 0.02 & 15.5\\
		2 & 0.041 & 1.90 $\pm$ 0.13 & 14.4 & 0.34 $\pm$ 0.04 & 7.7\\
        3 & 0.032 & 1.55 $\pm$ 0.20 & 7.6 & 0.35 $\pm$ 0.08 & 4.6\\
        4 & 0.025 & 1.33 $\pm$ 0.27 & 4.8 & 0.33 $\pm$ 0.07 & 4.5\\
		\hline
	\end{tabular}
\end{table}

\subsection{Aperture Size}
\label{sec:aperture}
\begin{figure}
    \centering
    \includegraphics[width=8cm]{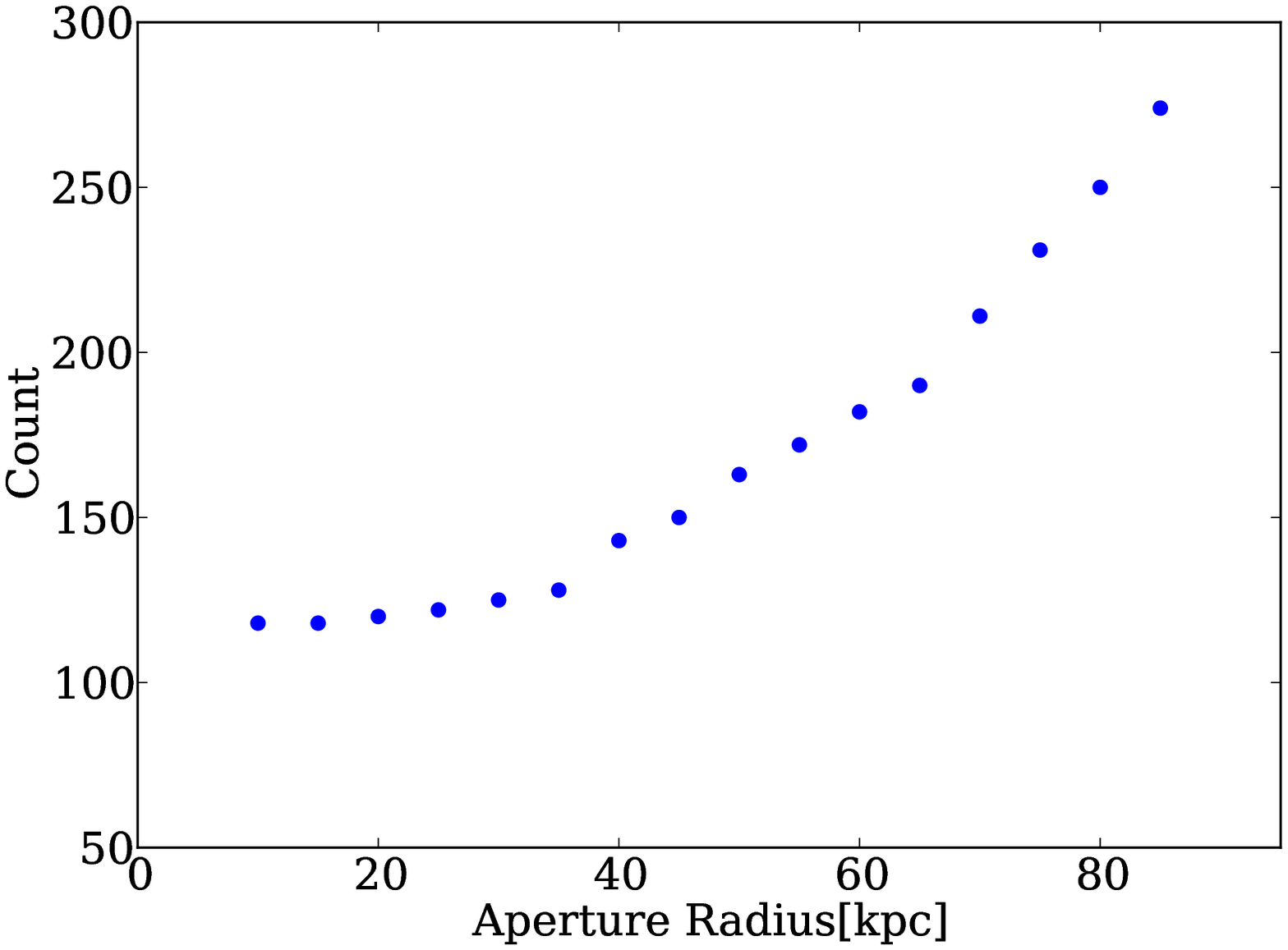}
    \includegraphics[width=8cm]
    {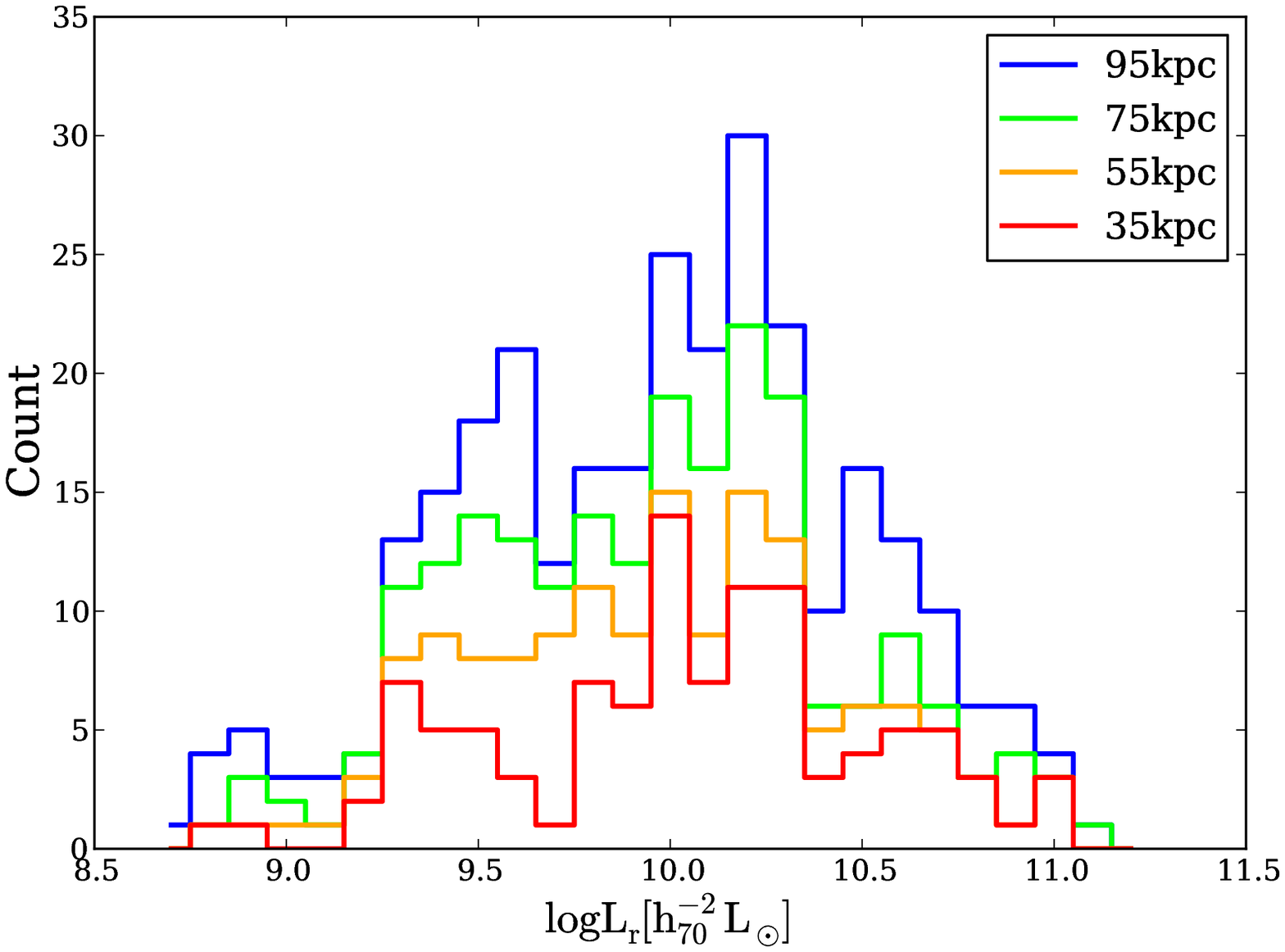}
    \caption{The top panel shows the number of confused galaxies in stacks of different aperture size. The bottom panel shows the histogram of the luminosity of the confused galaxies.}
    \label{aperture_statistics}
\end{figure}

With our relatively small synthesized beam area, many SDSS galaxies will be resolved or partially resolved in HI. The extraction radius therefore needs to be carefully chosen. Too small a radius may miss HI flux, while too large radius will unnecessarily introduce extra noise, and increase confusion from nearby galaxies. Based on determining the maximum radius prior to confusion becoming a problem (see Figure~\ref{aperture_statistics}), we have chosen an aperture radius of 35 kpc, similar to the 30 kpc box size used by \citet{2015A&A...580A..43G}, whose observations had a somewhat smaller $\sim 10\arcsec$ synthesized beam. 

Figure~\ref{aperture_statistics} shows that, for radii $<35$ kpc, the number of confused galaxies within (a) the aperture or within the synthesized beam, and (b) within 3 MHz (630 km s$^{-1}$) remains in the range 120 -- 130. However, at larger apertures, confusion increases rapidly, approximately doubling by 80 kpc. The luminosity distribution of the confused galaxies is shown in Figure~\ref{aperture_statistics}.

\begin{figure}
    \centering
    \includegraphics[width=8cm]{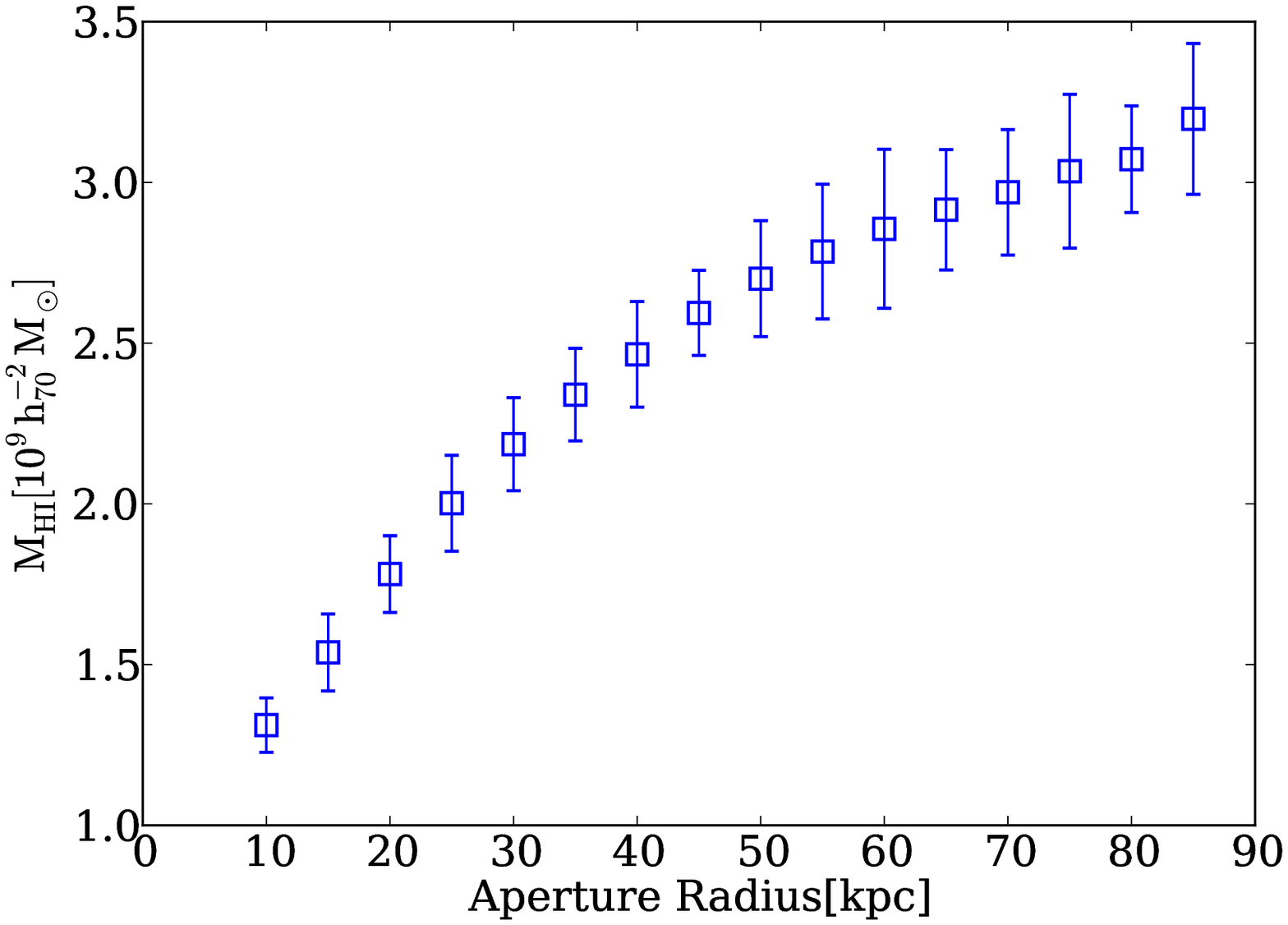}
    \includegraphics[width=8cm]{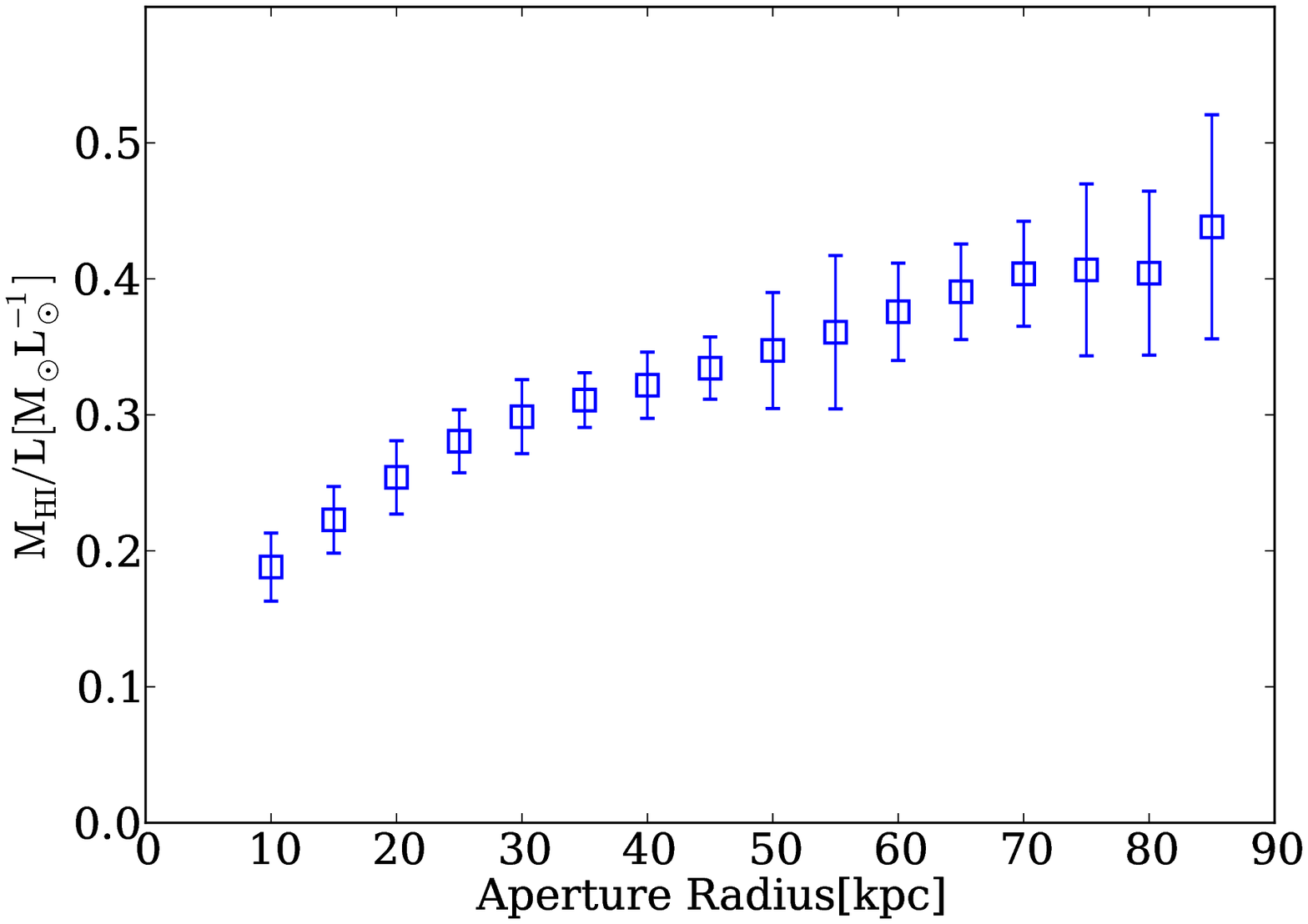}
    \caption{The measurements from stacks with different aperture size. The top panel is the for $M_{\rm HI}$ stacks and the bottom panel is for $M_{\rm HI}/L$ stacks.}
    \label{aperture_stacking}
\end{figure}

The corresponding stacked values for $\langle M_{\rm HI}\rangle$ and $\langle M_{\rm HI}/L\rangle$ are shown in Figure~\ref{aperture_stacking}. $\langle M_{\rm HI}\rangle$ increases monotonically, reflecting the finite size of the galaxy HI disks at small apertures, and the effect of confusion at large apertures. Between 35 and 80 kpc, $\langle M_{\rm HI}\rangle$ increases by 40 per cent. $\langle M_{\rm HI}/L\rangle$ is less sensitive to aperture. Values for both are given in Table~\ref{aperture_table}, and show that S/N ratio for $\langle M_{\rm HI}/L\rangle$ is maximized when the aperture radius is 35 kpc.

\subsection{Confusion correction}
\label{s:confusion}
As shown above, $\sim 7$ per cent of our sample is potentially confused with neighbouring galaxies, both catalogued and uncatalogued. Although the WSRT synthesized beam is almost an order of magnitude smaller than the Arecibo beam and two orders of magnitude smaller than the Parkes beam, we can nevertheless estimate the corresponding correction factors for $\langle M_{\rm HI}\rangle$ and $\langle M_{\rm HI}/L\rangle$. 

We have therefore carried out a mock stacking experiment on the TAIPAN+WALLABY simulations of \citet{2017PASA...34...47D}, who employ the state-of-the-art theoretical galaxy formation model {\sc GALFORM} \citep{2000MNRAS.319..168C}, in the version presented by \citet{2012MNRAS.426.2142L}. The latter follows the cosmic evolution of galaxies using a self-consistent two-phase interstellar medium model, in which stars form from the molecular gas content of galaxies. This model provides a physical distinction between atomic and molecular hydrogen in galaxies, and thus it is capable of predicting the evolution of these two components separately. The specific lightcones used here were produced using the $N$-body cold dark matter cosmological Millennium I \citep{2005Natur.435..629S} and II \citep{2009MNRAS.398.1150B} simulations, which in combination allow us to have a complete census of the HI masses of galaxies from the most HI-massive galaxies, down to an HI mass of $\approx 10^6$ M$_{\odot}$. Two sets of lightcones were created and presented in \citet{2017PASA...34...47D}, one mimicking the selection function of TAIPAN and another one mimicking the selection function of WALLABY, with the primary aim of assessing the overlap population between the two surveys. Here, we use only the WALLABY\footnote{This is the Extragalactic All Sky HI Survey being carried out with the Australian Square Kilometer Array Pathfinder \citep{2008ExA....22..151J}.} lightcones.

We extract 100 strips each of $2\fdg5 \times 40^{\circ}$, and in each strip we produce 35 pointings of radii 0\fdg5. We select the galaxies located in these 35 pointings from $z = 0.01\sim 0.11$ and $r\le17.7$. We also produce a volume-limited sub-sample as previously described. We use the same method as above to measure the $\langle M_{\rm HI}\rangle$ and $\langle M_{\rm HI}/L\rangle$, after locating the confused galaxies. We carry out the stacking using three methods:
\begin{enumerate}
\item Assuming that there is no confusion (i.e.\ stack the HI in the selected galaxies only);
\item Combine the HI in the sample galaxies with that of any companions with $r<17.77$;
\item Combine the HI in the sample galaxies with that of all companions. 
\end{enumerate}

We follow the method in \citet{2012MNRAS.427.2841F} to model the confusion, estimating the total signal $S_{i}$ as the sum of the sample galaxy $S_{s}$ and the companions ($S_{c}$) weighted with two factors:
\begin{eqnarray}
     S_{i} = S_{s} + \Sigma_{c}f_{1;c}f_{2:c}S_{c},
    \label{signal_with_confusion}
\end{eqnarray}
where the $f_{1}$ and $f_{2}$ model the overlap between the sample galaxy and its companion in angular and redshift space.

The results are shown in Table~\ref{confusion_simulation}. For the magnitude-limited sample, the value of $\langle M_{\rm HI}\rangle$ derived from stacking confused sample galaxies with $r \le 17.77$ and stacking with all confused galaxies, are 1.3 $\pm$ 0.6 and 2.1 $\pm$ 0.7 per cent larger than the `correct' result, respectively. For $\langle M_{\rm HI}/L\rangle$, the increase is 1.4 $\pm$ 0.6 and 1.7 $\pm$ 0.6 per cent, respectively. The increments for the volume-limited sub-samples are slightly more. For the real data, we will later utilise the confusion-included sample and use correction factors based on the ratios of method (i) and (iii) above, with 35kpc resolution.

\begin{table}
    \centering
    \caption{Average HI mass $M_{\rm HI}$, mass-to-light ratio $M_{\rm HI}/L$, and corresponding signal-to-noise ratio $S/N$ as a function of aperture size.}
    \label{aperture_table}
    \begin{tabular}{cccccc}
        \hline
        Aperture & <$\rm M_{HI}$> & $\rm S/N(M_{HI})$ & <$\rm M_{HI}/L$> & $\rm S/N(M_{HI}/L)$\\
        (kpc)&($10^{9}h_{70}^{-2}M_{\odot}$)& &($M_{\odot}L_{\odot}^{-1}$)& \\
        \hline
        10 & 1.31 $\pm$ 0.08 & 15.5 & 0.19 $\pm$ 0.03 & 7.5\\
        15 & 1.54 $\pm$ 0.12 & 12.9 & 0.22 $\pm$ 0.02 & 9.1\\
        20 & 1.78 $\pm$ 0.12 & 14.9 & 0.25 $\pm$ 0.03 & 9.4\\
        25 & 2.00 $\pm$ 0.15 & 13.4 & 0.28 $\pm$ 0.02 & 12.1\\
        30 & 2.18 $\pm$ 0.14 & 15.1 & 0.30 $\pm$ 0.03 & 11.0\\
        35 & 2.34 $\pm$ 0.14 & 16.3 & 0.31 $\pm$ 0.02 & 15.5\\
        40 & 2.46 $\pm$ 0.16 & 15.0 & 0.32 $\pm$ 0.02 & 13.2\\
        45 & 2.59 $\pm$ 0.13 & 19.6 & 0.33 $\pm$ 0.02 & 14.6\\
        50 & 2.70 $\pm$ 0.18 & 15.0 & 0.35 $\pm$ 0.04 & 8.1\\
        55 & 2.78 $\pm$ 0.21 & 13.3 & 0.36 $\pm$ 0.06 & 6.4\\
        60 & 2.86 $\pm$ 0.25 & 11.6 & 0.38 $\pm$ 0.04 & 10.5\\
        65 & 2.91 $\pm$ 0.19 & 15.6 & 0.39 $\pm$ 0.04 & 11.1\\
        70 & 2.97 $\pm$ 0.20 & 15.2 & 0.40 $\pm$ 0.04 & 10.5\\
        75 & 3.03 $\pm$ 0.24 & 12.7 & 0.41 $\pm$ 0.06 & 6.4\\
        80 & 3.07 $\pm$ 0.17 & 18.5 & 0.40 $\pm$ 0.06 & 6.7\\
        85 & 3.20 $\pm$ 0.23 & 13.6 & 0.44 $\pm$ 0.08 & 5.3\\
		\hline
	\end{tabular}
\end{table}

\begin{table}
	\centering
	\caption{Measurements of $\langle M_{\rm HI}\rangle$ and $\langle M_{\rm HI}/L\rangle$ from stacking galaxies in the mock catalog. The superscript `m' and `v' refer to the magnitude-limited sample and volume-limited sub-sample respectively.}
	\label{confusion_simulation}
	\begin{tabular}{cccc} 
		\hline
		 & No confusion & Confused with   & Confused with \\
         &              & sample galaxies & all galaxies  \\
		\hline
		$\langle M^{\rm m}_{\rm HI}\rangle(10^{9}h_{70}^{-2}$ M$_{\odot}$) & 2.757  & 2.794  & 2.816\\
		$\langle M^{\rm m}_{\rm HI}/L\rangle$(M$_{\odot}/$L$_{\odot}$) & 0.289  & 0.293 & 0.294\\
        $\langle M^{\rm v}_{\rm HI}\rangle(10^{9}h_{70}^{-2}$ M$_{\odot}$) & 1.674 & 1.706 & 1.708\\
		$\langle M^{\rm v}_{\rm HI}/L\rangle$(M$_{\odot}/$L$_{\odot}$) & 0.560 & 0.570 & 0.572 \\

		\hline
	\end{tabular}
\end{table}

\subsection{PSF effects}
\begin{figure*}
    \centering
    \includegraphics[width=16cm]{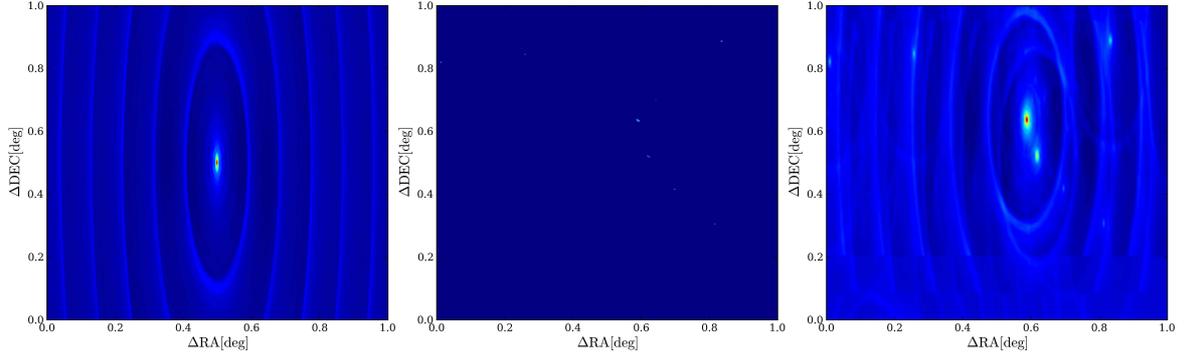}
    \caption{Left panel: average PSF of WSRT observations. Central panel: a subset of a 3 MHz slice of the $S^3$-SAX simulated sky at $z=0.055$. Right panel: the same slice after the convolution with the PSF.}
    \label{beam_convolve}
\end{figure*}

In interferometric observations, the original HI sky is convolved with the point spread function (PSF) of the telescope. The PSF is then normally removed using a deconvolution algorithm. However, such a procedure is not possible when individual galaxy signals are below the noise level. Our stacks are therefore stacks of `dirty' maps. To explore the effect of this, we again employ a simulation. 

We convolve a simulated HI sky with the average PSF of WSRT.  The simulated sky is based on the mock catalogue of \citet{2012MNRAS.426.3385D}, in which the Theoretical Astrophysical Observatory (TAO) was used to generate a light-cone catalogue from the semi-analytic models of \citet{2006MNRAS.365...11C}.  Cold gas masses in this simulation were scaled by Duffy et al. to match the local HI mass function measured by ALFALFA \citep{2010ApJ...723.1359M} to ensure a realistic modelling of the local HI Universe. Galaxies with HI masses $M_{\rm  HI} > 10^{8.5}$ or apparent magnitudes $m_r < 19.8$ are populated into the synthetic sky using the GALMOD routine from GIPSY.

In Figure~\ref{beam_convolve}, we illustrate the convolution process. The left panel is the PSF of WSRT, the central panel shows a 3 MHz slice of the simulated sky at $z=0.055$, and the right panel is the same slice after the convolution with the PSF. We can see clearly see the effect of sidelobes on the surrounding sky. To quantify this effect, we apply the same stacking method to the simulated sky and the convolved sky. We stack the spectra from 2727 galaxies located in the range $0.01 < z < 0.11$ with apparent $r$-band magnitudes brighter than 17.7. 

In Table~\ref{convolve_table}, we show the results of stacking with the original catalogue, the simulated sky and the convolved sky. For the latter two, we use an aperture with a radius of 35 kpc to extract the spectra. Directly stacking the HI mass given by the catalogue results in an averaged HI mass of 3.013 $\times10^{9} h_{70}^{-2}$ M$_{\odot}$. Stacking the spectra of the selected galaxies in the simulated sky gives 3.021 $\times10^{9} h_{70}^{-2}$ M$_{\odot}$, higher due to confusion. Stacking the spectra obtained from the convolved sky gives 2.962 $\times10^{9} h_{70}^{-2}$ M$_{\odot}$, lower due to the inclusion of negative sidelobes. Convolution makes the averaged integrated flux smaller by $1.7\%$, meaning that sidelobes only result in a small underestimate of the true signal. 

\begin{table}
	\centering
	\caption{The results of stacking with the original catalogue, the simulated sky and the convolved sky. For the latter two, we use a aperture with a radius of 35 kpc to extract the spectra.}
	\label{convolve_table}
	\begin{tabular}{lcc} 
		\hline
		Data source & Aperture & Stacked integral\\
             &    (kpc)      & ($10^9h_{70}^{-2}$ M$_{\odot}$) \\
		\hline
		$S^3$-SAX catalogue & -- & 3.013\\
		Confused sky & 35 & 3.021\\
		Convolved sky & 35 & 2.962\\
		\hline
	\end{tabular}
\end{table}

\subsection{Cosmic Variance}
The Universe is only homogeneous on scales $>> 100$ Mpc \citep{2012MNRAS.425..116S}. Therefore observations in smaller regions can be affected by small-scale inhomogeneity, or cosmic variance. To assess the effect on our results, we assume the WSRT pointings are conical and we assign the `beam edge' as the radius at which the normalized primary beam response equals to 0.1. At the median redshift of 0.066, the radius of this beam $r_{z=0.066,f=0.1}$ = 0.5195 deg, corresponding to 2368 kpc. This corresponds to a comoving volume of 6642 Mpc$^{3}$ per pointing with the small volume at $z < 0.01$ removed. The weighted noise-equivalent volume (square primary beam weighting) for each beam is 1545 Mpc$^{3}$. The number of SDSS galaxies with spectroscopic redshifts in each pointing varies between 18 and 146 (see Table~\ref{stacking_pointings}). Combining the 35 pointings together, the weighted sampled volume is $5.4\times 10^{4}$ Mpc$^{3}$, which can be compared with the sampled volumes of HIPASS ($2.37\times 10^{5}$ Mpc$^{3}$, \citet{2005MNRAS.359L..30Z}) and the 100 per cent ALFALFA source catalog ($10.15\times 10^{5}$ Mpc$^{3}$, \citet{2018MNRAS.tmp..502J}).

A simple quantifiable measure of the cosmic variance can be obtained by examining the variance of galaxy counts in the TAIPAN+WALLABY simulation. We define $\xi[\%] = 100\times \sigma_{\rm cv}/\langle N\rangle$, where the variance $\sigma^{2}_{\rm cv} = \Sigma_{i}(\langle N\rangle-N_{i})^2/n$, $\langle N\rangle$ is the mean galaxy count in the selected volumes, $N_{i}$ the number of galaxies in the volume $i$ and $n$ the total number of selected volumes. We randomly select 1000 strips of the same size as the WSRT strip and with the same redshift region from the simulation. In each strip we produce 35 pointings whose distributions are same as the WSRT observations. For galaxies within 0\fdg5 of one of the pointing centres, we find $\xi = 9.1\pm 0.3$ per cent. 

For SDSS in the main region, the mean weighted number of galaxies at Declinations near $14^{\circ}$ across a similar 35 simulated pointings is $465$ (reduced from $1485$ by primary beam weighting), with a similar cosmic variance of 12\%. However, the weighted number of galaxies in our sample is substantially higher at 519 (reduced from 1895 by weighting).
This implies that the region observed is overdense by more than the variation expected from cosmic variance.  
Nevertheless, the cosmic variance across a wide field of view is much lower compared with a deep single pointing. Furthermore, normalization using the SDSS luminosity function removes first-order changes to the HI density associated with optical overdensities. However, second-order environmental effects may influence the final result.

\section{Results}
\label{sec:results}
\subsection{Individual Pointings}
\label{sec:samples}
The magnitude-limited sample has a mean redshift of $\langle z \rangle = 0.066$. The stacking results for each individual pointing are given in Table~\ref{stacking_pointings}. Because of fewer galaxies and a smaller effective volume, the errors (estimated with jackknife method) are larger compared with the results from stacking the total sample. For the stacked mass spectra, only one stack (pointing 17) does not show a detection, three (pointings 12, 29 and 35) have unclear detections, while the remaining 30 pointings all result in clear detections. We show the stacked mass spectra in Appendix~\ref{sec:StackedSpectra}.

\begin{table*}
    \caption{Results from stacking HI spectra in the 35 individual pointings observed with the WRST. Because of the smaller sample, the effective volume reduces and cosmic variance increases, the errors are larger compared with the results from the stacking with total sample(section~\ref{sec:weight}). We also show the statistical errors for stacked mass and mass-to-light spectra as noise$_{\rm stat,m}$ and noise$_{\rm stat,m/l}$.}
    \label{stacking_pointings}
    \begin{tabular}{ccccccccc}
        \hline
       Pointing & Position & $\rm N$ & <$\rm z$> & <$\rm M_{\rm HI}$> & noise$_{\rm stat,m}$ & <$\rm M_{\rm HI}/L$> & noise$_{\rm stat,m/l}$ & Obs time \\
                 & (J2000) &     &      & ($10^{9}h_{70}^{-2}$ M$_{\odot}$) & ($10^{9}h_{70}^{-2}$ M$_{\odot}$) & (M$_{\odot}/$L$_{\odot}$) & (M$_{\odot}/$L$_{\odot}$) &  (hrs) \\
        \hline
       1 & 22:27:00 +13:37:48 & 36 & 0.092  & 5.15 $\pm$ 3.96 & 0.41 & 1.33 $\pm$ 1.11 & 0.11 & 12.0\\
       2 & 22:37:48 +14:18:36 & 66 & 0.074  & 3.43 $\pm$ 0.57 & 0.24 & 0.22 $\pm$ 0.08 & 0.03 & 12.0\\
       3 & 22:57:50 +13:03:36 & 45 & 0.057  & 1.80 $\pm$ 0.53 & 0.11 & 0.70 $\pm$ 0.31 & 0.04 & 12.0\\
       4 & 23:12:58 +13:56:24 & 71 & 0.066  & 5.38 $\pm$ 0.84 & 0.20 & 0.75 $\pm$ 0.21 & 0.04 & 11.5\\
       5 & 23:14:24 +14:39:00 & 49 & 0.074  & 4.94 $\pm$ 1.42 & 0.30 & 0.45 $\pm$ 0.10 & 0.03 & 11.0\\
       6 & 23:24:54 +15:18:00 & 70 & 0.056  & 3.18 $\pm$ 0.55 & 0.11 & 0.35 $\pm$ 0.10 & 0.02 & 10.7\\
       7 & 23:43:23 +14:16:08 & 36 & 0.073  & 11.13 $\pm$ 4.46 & 0.59 & 0.76 $\pm$ 0.37 & 0.05 & 9.8\\
       8 & 23:51:36 +14:06:00 & 46 & 0.078  & 5.17 $\pm$ 1.15 & 0.27 & 0.53 $\pm$ 0.10 & 0.04 & 8.8\\
       9 & 02:03:18 +13:51:00 & 31 & 0.063  & 2.28 $\pm$ 1.18 & 0.31 & 0.18 $\pm$ 0.08 & 0.03 & 11.3\\
       10 & 22:12:29 +12:20:24 & 31 & 0.067 &  1.44 $\pm$ 0.48 & 0.17 & 0.44 $\pm$ 0.17 & 0.04 & 12.0\\
       11 & 22:14:38 +13:52:12 & 81 & 0.044 &  0.33 $\pm$ 0.14 & 0.05 & -- & -- & 9.7\\
       12 & 22:33:18 +13:11:02 & 35 & 0.089 &  3.98 $\pm$ 1.41 & 0.62 & 0.32 $\pm$ 0.09 & 0.04 & 12.0\\
       13 & 22:39:00 +13:26:24 & 57 & 0.079 &  1.58 $\pm$ 1.07 & 0.28 & 0.60 $\pm$ 0.74 & 0.06 & 12.0\\
       14 & 23:18:18 +14:55:12 & 39 & 0.081 &  3.41 $\pm$ 1.22 & 0.44 & 0.30 $\pm$ 0.10 & 0.08 & 10.7\\
       15 & 23:26:24 +14:03:00 & 55 & 0.054 &  2.11 $\pm$ 0.49 & 0.16 & 0.43 $\pm$ 0.12 & 0.05 & 10.3\\
       16 & 23:38:06 +15:45:43 & 60 & 0.066 &  2.53 $\pm$ 0.82 & 0.21 & 0.19 $\pm$ 0.08 & 0.03 & 8.6\\
       17 & 23:45:36 +15:22:12 & 26 & 0.087 &  -- & -- & 0.06 $\pm$ 0.21 & 0.08 & 9.3\\
       18 & 23:56:53 +13:57:00 & 27 & 0.067 &  8.92 $\pm$ 1.93 & 0.29 & 0.70 $\pm$ 0.14 & 0.04 & 12.0\\
       19 & 00:00:36 +15:24:36 & 28 & 0.077 &  1.60 $\pm$ 3.55 & 0.38 & 0.05 $\pm$ 0.16 & 0.02 & 5.4\\
       20 & 00:06:00 +15:43:48 & 36 & 0.069 &  5.91 $\pm$ 2.71 & 0.29 & 0.32 $\pm$ 0.19 & 0.03 & 10.0\\
       21 & 00:24:00 +14:12:00 & 18 & 0.060 &  2.03 $\pm$ 1.79 & 0.21 & 0.15 $\pm$ 0.34 & 0.06 & 6.1\\
       22 & 00:43:01 +15:18:00 & 74 & 0.080 &  1.34 $\pm$ 0.71 & 0.17 & 0.10 $\pm$ 0.04 & 0.01 & 10.8\\
       23 & 01:10:03 +13:59:49 & 91 & 0.061 &  0.22 $\pm$ 0.68 & 0.09 & 0.07 $\pm$ 0.09 & 0.01 & 12.0\\
       24 & 01:11:28 +15:06:00 & 63 & 0.055 &  3.96 $\pm$ 1.28 & 0.20 & 0.47 $\pm$ 0.14 & 0.03 & 12.0\\
       25 & 01:15:00 +14:28:48 & 66 & 0.064 &  3.94 $\pm$ 0.60 & 0.23 & 0.46 $\pm$ 0.11 & 0.03 & 4.6\\
       26 & 01:55:48 +14:45:07 & 75 & 0.068 &  3.10 $\pm$ 0.69 & 0.23 & 0.64 $\pm$ 0.17 & 0.03 & 10.1\\
       27 & 01:57:11 +13:09:00 & 51 & 0.057 &  3.93 $\pm$ 0.97 & 0.20 & 0.43 $\pm$ 0.09 & 0.04 & 5.7\\
       28 & 02:12:00 +14:02:24 & 40 & 0.048 &  2.53 $\pm$ 0.84 & 0.11 & 0.84 $\pm$ 0.43 & 0.04 & 10.1\\
       29 & 00:00:36 +14:33:00 & 64 & 0.086 &  1.89 $\pm$ 1.43 & 0.43 & 0.18 $\pm$ 0.08 & 0.03 & 9.4\\
       30 & 00:30:36 +14:52:12 & 33 & 0.074 &  2.29 $\pm$ 1.68 & 0.37 & 0.09 $\pm$ 0.13 & 0.04 & 7.8\\
       31 & 00:58:01 +14:50:24 & 54 & 0.074 &  4.21 $\pm$ 1.19 & 0.35 & 0.26 $\pm$ 0.08 & 0.03 & 9.5\\
       32 & 01:19:48 +14:45:40 & 57 & 0.050 &  1.43 $\pm$ 0.36 & 0.14 & 0.36 $\pm$ 0.24 & 0.03 & 9.8\\
       33 & 01:46:30 +13:51:00 & 42 & 0.062 &  2.88 $\pm$ 1.04 & 0.25 & 0.35 $\pm$ 0.13 & 0.04 & 12.0\\
       34 & 01:49:26 +13:51:00 & 88 & 0.062 &  3.37 $\pm$ 0.52 & 0.14 & 0.41 $\pm$ 0.08 & 0.03 & 10.8\\
       35 & 23:24:18 +14:40:48 & 154 & 0.052 &  0.51 $\pm$ 0.30 & 0.11 & 0.08 $\pm$ 0.05 & 0.03 & 9.4\\
        \hline
    \end{tabular}
\end{table*}

\subsection{All Galaxies}
\begin{figure}
    \centering
    \includegraphics[width=8cm]{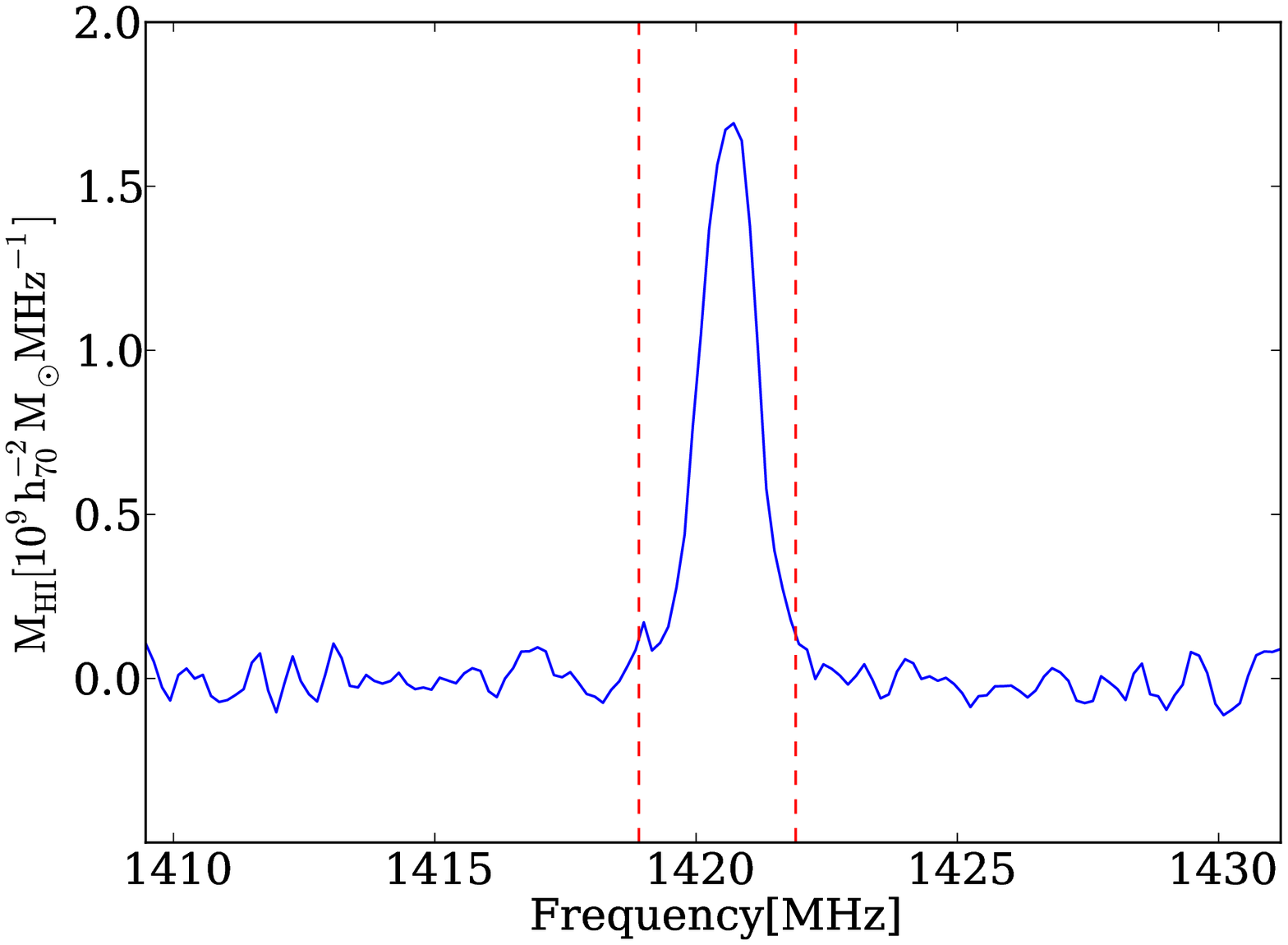}
    \includegraphics[width=8cm]{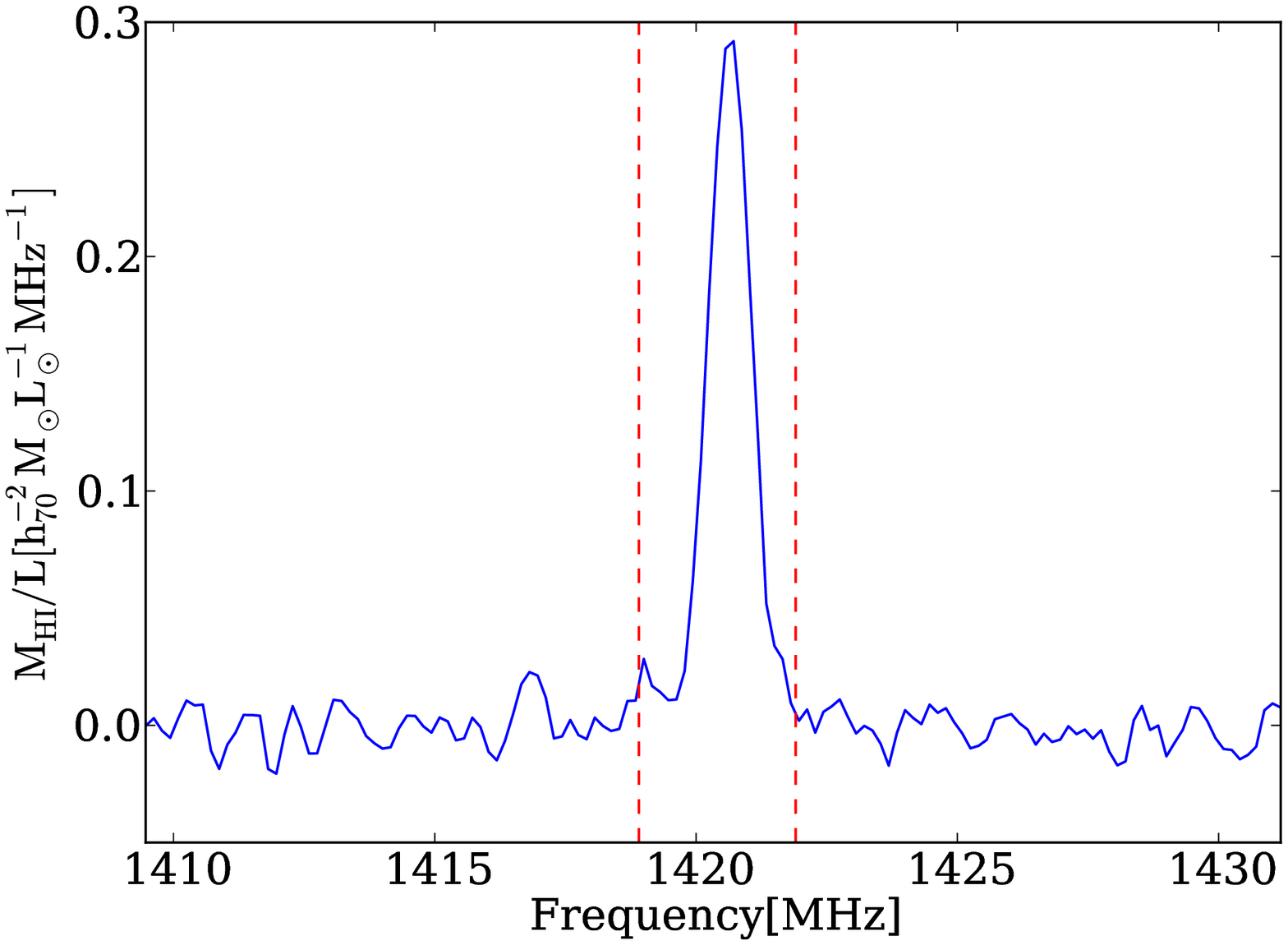}
    \caption{Stack of all galaxies in the magnitude-limited sample. Top plot: stack of the mass spectra showing a clear 67$\sigma$ detection. Integrating the spectral line and applying the confusion correction results in an HI mass of $M_{\rm HI} = (2.29 \pm 0.13)\times 10^{9} h_{70}^{-2}$ M$_{\odot}$. Bottom plot: stack of mass-to-light ratio spectra resulting in a clear 56$\sigma$ detection with $\langle M_{\rm HI}/L\rangle = (0.306 \pm 0.020)$ M$_{\odot}/$L$_{\odot}$. The red-dashed line indicates the region of the integration.}
    \label{total_stacking}
\end{figure}

Stacking all mass spectra from our magnitude-limited sample results in a strong 67$\sigma$ detection, where the noise level is estimated from the jackknife sampling. We measured the HI mass of the stack in the manner described in Section~\ref{sec:script}. Integrating the spectral line over the rest frequency range of $\nu = 1420.406 \pm 1.5$ MHz and applying the confusion correction results in a mean HI mass $\langle M_{\rm HI}\rangle = (2.29 \pm 0.13)\times 10^9 h_{70}^{-2}$ M$_{\odot}$. The mean stacked value for the ratio $\langle M_{\rm HI}/L\rangle$ ratio results in a 56$\sigma$ detection with $\langle M_{\rm HI}/L\rangle = (0.306\pm 0.020)$ M$_{\odot}/$L$_{\odot}$. The stacked spectra are shown in Figure~\ref{total_stacking}. For the volume-limited sub-sample, we obtain $\langle M_{\rm HI}\rangle = (0.844 \pm 0.129)\times 10^{9} h_{70}^{-2}$ M$_{\odot}$ and $\langle M_{\rm HI}/L\rangle = (0.369 \pm 0.095)$ M$_{\odot}/$L$_{\odot}$. 

\subsection{Redshift Bins}
\label{sec:redshiftbins}
\begin{figure}
    \centering
    \includegraphics[width=8cm]{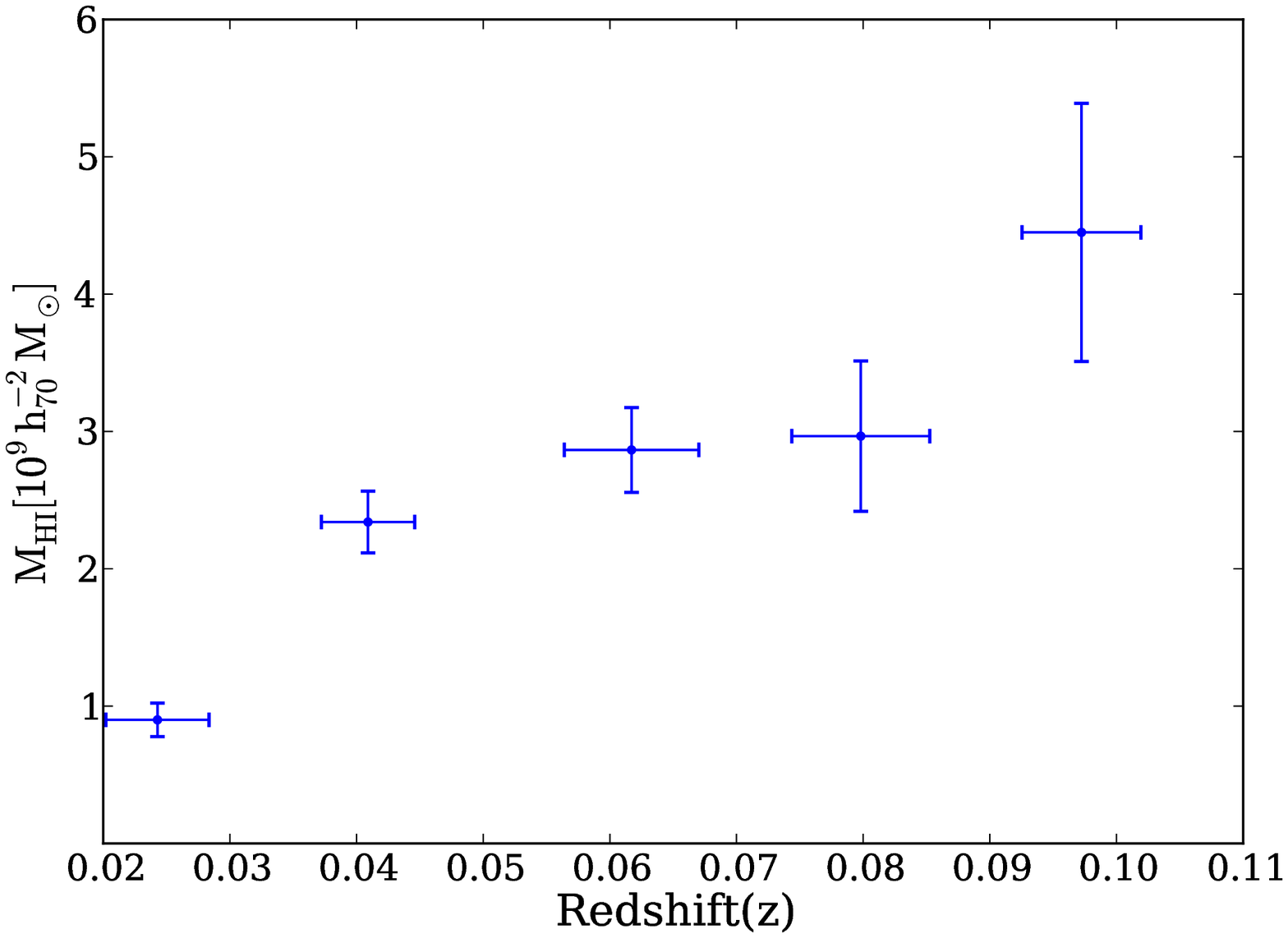}
    \includegraphics[width=8cm]{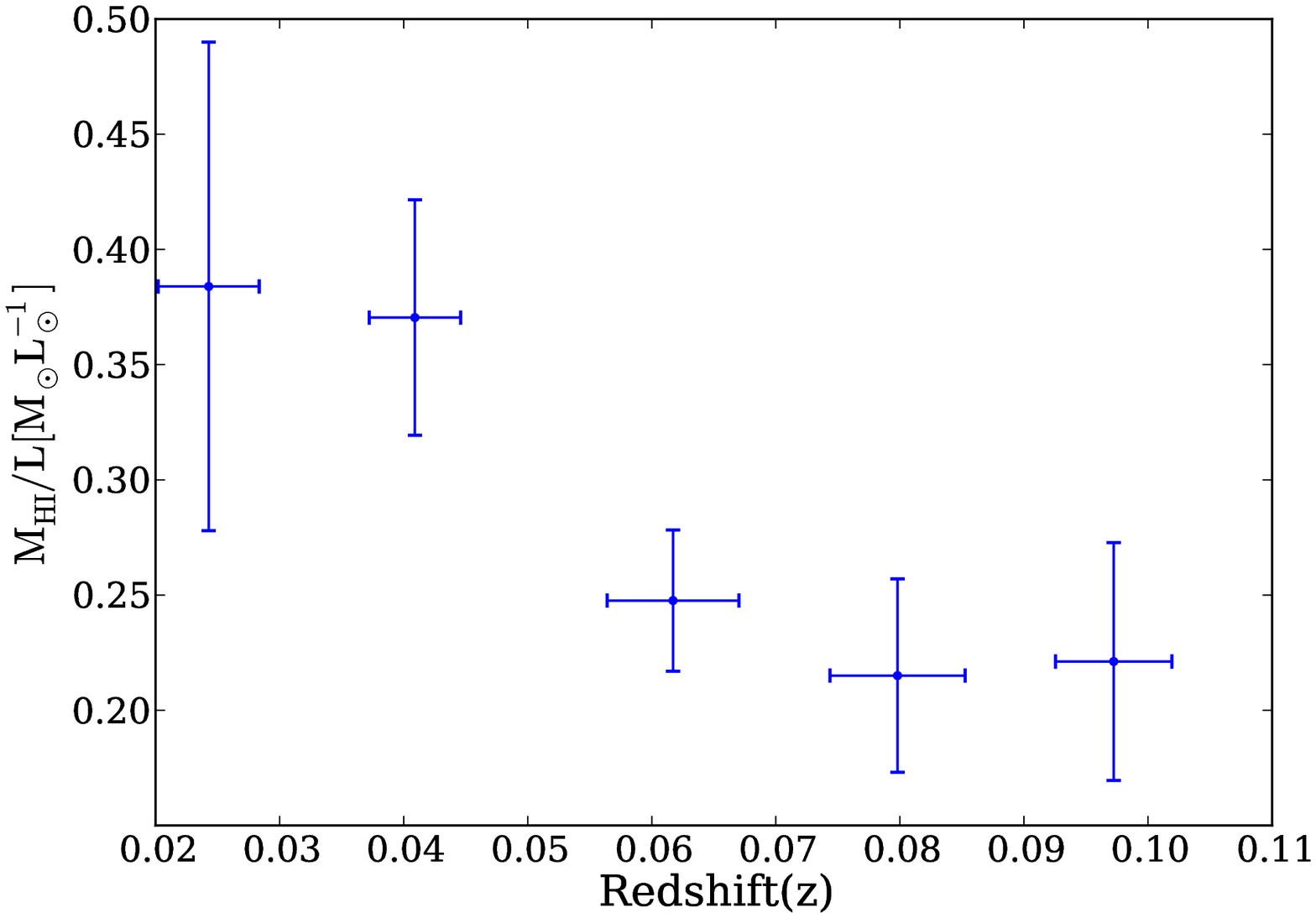}
    \caption{The results of stacking in redshift bins.  The top panel is for HI mass $M_{\rm HI}$, and the bottom panel is for $M_{\rm HI}/L$. The data points are centred at the mean redshift of each bin. The redshift error bars represent the 1$\sigma$ standard deviation within each bin.}
    \label{redshift_bins}
\end{figure}

The large redshift region and selection effects results in the sample properties changing with redshift. We split the sample into five redshift bins. The mean redshift of each bin is $\langle z \rangle =$ 0.024, 0.041, 0.062, 0.080 and 0.097. The sub-samples contain 155, 439, 453, 448 and 400 galaxies, respectively. All stacks result in significant detections. The derived average HI masses $\langle M_{\rm HI}\rangle$ and HI mass-to-light ratios $\langle M_{\rm HI}/L_{\rm r}\rangle$ are shown in Figure~\ref{redshift_bins} and Table~\ref{stacking_redshift_table}. The HI mass increases with redshift, and $M_{\rm HI}/L_{\rm r}$ decreases. Both results are explained by the fact that the samples are biased towards more luminous galaxies at higher redshift (see Figure~\ref{stacking_lumibins}). 

\begin{table}
	\centering
	\caption{The results of stacking the sample in different redshift bins.}
	\label{stacking_redshift_table}
	\begin{tabular}{cccccc} 
		\hline
		$\langle z\rangle$ & redshift range & $N_{\rm g}$ & <$\rm M_{\rm HI}$> & <$\rm M_{\rm HI}/L$>\\
        & & &($10^{9}h_{70}^{-2}$M$_{\odot}$) & (M$_{\odot}/$L$_{\odot}$) \\
		\hline
		0.024 $\pm$ 0.004 & 0.01 - 0.03 & 155 & 0.90 $\pm$ 0.12 & 0.38 $\pm$ 0.11\\
		0.041 $\pm$ 0.004 & 0.03 - 0.05 & 439 & 2.34 $\pm$ 0.22 & 0.37 $\pm$ 0.05\\
		0.062 $\pm$ 0.005 & 0.05 - 0.07 & 453 & 2.87 $\pm$ 0.31 & 0.25 $\pm$ 0.03\\
        0.080 $\pm$ 0.005 & 0.07 - 0.09 & 448 & 2.97 $\pm$ 0.55 & 0.21 $\pm$ 0.04\\
        0.097 $\pm$ 0.005 & 0.09 - 0.11 & 400 & 4.45 $\pm$ 0.94 & 0.22 $\pm$ 0.05\\
        \hline
	\end{tabular}
\end{table}

\section{Cosmic HI Density $\Omega_{\rm HI}$}
\label{sec:density}
\subsection{Luminosity Bias}
\label{sec:magnitudebias}
\begin{figure} 
    \centering
    \includegraphics[width=8cm]{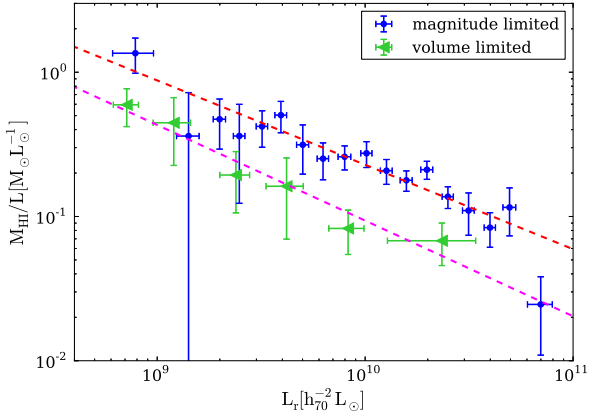}
    \caption{Stacking the magnitude-limited (blue circle) and volume-limited (green triangle) sample in luminosity bins shows that $\rm M_{HI}/L$ decreases with increasing luminosity. The red dashed line indicates the best-fit to the magnitude-limited sample of  $\log(M_{\rm HI}/L) = (-0.587 \pm 0.046)\log L + (5.246 \pm 0.517)$; the magenta dashed line shows the best-fit to the volume-limited data of $\log(M_{\rm HI}/L) = (-0.662 \pm 0.120)\log L + (5.600 \pm 1.068)$.}
    \label{stacking_lumibins}
\end{figure}

\begin{figure}
    \centering
    \includegraphics[width=8cm]{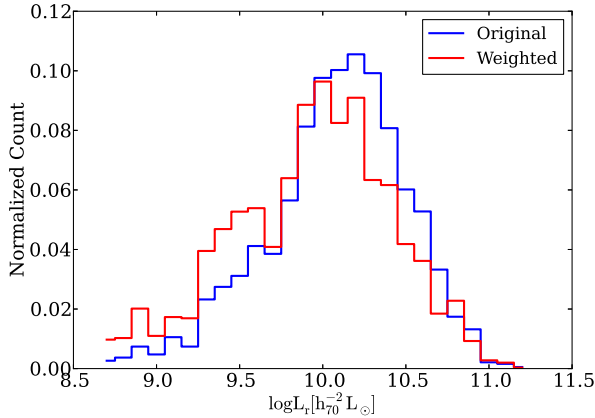}
    \caption{The distribution of SDSS galaxies from the magnitude-limited sample in $r$-band luminosity bins. The interval is 0.1 dex. The blue line represents the luminosities of galaxies in the sample; the red line shows the weighted number of the galaxies in the same luminosity bins.}
    \label{luminosity_distribution}
\end{figure}

SDSS is a magnitude-limited sample and therefore many optically faint, but HI-rich galaxies are missed at higher redshift (Figure~\ref{redshift-L_r}). This has an influence on our results for $M_{\rm HI}$ and $M_{\rm HI}/L$ and means that we sample different populations of galaxies at different redshifts. To account for the missed faint, but high $M_{\rm HI}/L$ ratio galaxies, we assume a power-law relation between $M_{\rm HI}/L$ and luminosity given by $M_{\rm HI}/L \sim L^{\beta}$. $\beta$ is obtained from stacking galaxies binned by their $r$-band luminosity. We show the results in Figure~\ref{stacking_lumibins}. There is a significant decrease of $M_{\rm HI}/L$ with increasing $L_r$. We find $\log(M_{\rm HI}/L) = (-0.587 \pm 0.046)\log L + (5.246 \pm 0.517)$. 
Since the sample is not complete in $r$-band luminosity at all redshifts, there is a selection effect in favour of low values of $M_{\rm HI}/L$ and high values of $L$ in this plot. However, only the slope of this line is relevant for the current purposes and  the result appears to be similar to that derived from our the volume-limited sub-sample ($\beta=-0.662 \pm 0.120$ -- also shown in Figure~\ref{stacking_lumibins}). With this relation, a suitable correction for the $\langle M_{\rm HI}/L\rangle$ ratio is then given by \citet{2013MNRAS.433.1398D}: 
\begin{eqnarray}
     C1 = \frac{\langle M_{\rm HI}/L \rangle _{\rm all}}{\langle M_{\rm HI}/L \rangle _{\rm obs}}
            = \frac{\sum_{i=1}^{N} w_{i}}{\sum_{i=1}^{N} w_{i} (L_{i}/L_{\ast})^{\beta}}\times\frac{\int (L/L_{\ast})^{\beta}L\phi_{L}(L)dL}{\int L\phi_{L}(L)dL},
    \label{C1}
\end{eqnarray}
where $\phi_{L}(L)$ is the luminosity function, $w_{i}$ is the weight of i-th galaxy and $N$ = 1895. We use $L_{\ast}$ and $\phi_{L}(L)$ given by \citet{2003ApJ...592..819B}, where $\phi_{L}(L)$ is a Schechter function of the form:
\begin{eqnarray}
     \phi_{L}(L)dL = \phi^{\ast}\left(\frac{L}{L_{\ast}}\right)^{\alpha}\exp\left(-\frac{L}{L_{\ast}}\right)\frac{dL}{L_{\ast}},
    \label{schechter}
\end{eqnarray}
with the following parameters: $\phi^{\ast} = 5.11\times10^{-3}h_{70}^3$ Mpc$^{-3}$, $\log (L_{\ast}/L_{\odot})=10.36+\log h_{70}$ and $\alpha= -1.05$. Figure~\ref{luminosity_distribution} shows the original and weight-corrected distribution of SDSS galaxies in $r$-band luminosity bins. The weight shifts the original distribution to lower-luminosity bins because nearby galaxies are given more weight than distant galaxies (most of which are bright). We find a correction factor of C1 = 1.38.

\subsection{Stacked measurement of $\Omega_{\rm HI}$}
\label{sec:omegahi}
We calculate the cosmic HI density $\rho_{\rm HI}$ from the $\langle M_{\rm HI}/L\rangle$ ratio of the stack and the luminosity density derived for SDSS galaxies. The luminosity density for $z = 0.1$ in the $r$-band is given by $\rho_{L} = 1.29 \times10^8 h_{70}$ L$_{\odot}$ Mpc$^{-3}$ \citep{2003ApJ...592..819B} using 147,986 galaxies. Together with the correction factor C1, the HI density can be calculated according to:
\begin{eqnarray}
     \rho_{\rm HI} = C1\times\langle\frac{M_{\rm HI}}{L}\rangle\times\rho_{L}.
     \label{critical_HI_density}
\end{eqnarray}

We then correct confusion according to the method described in Section~\ref{s:confusion}. The correction for $\Omega_{\rm HI}$ is 1.7 $\pm$ 0.6 percent. Binning the galaxies into three redshift bins gives similar factors: 1.013 $\pm$ 0.006, 1.013 $\pm$ 0.006 and 1.038 $\pm$ 0.010 at mean redshifts of $\langle z\rangle$ = 0.038, 0.067 and 0.093, respectively. 

After applying the above corrections for luminosity bias and confusion, we calculate a local density of $\rho_{\rm HI} = (5.46 \pm 0.36)\times10^{7}h_{70}$ M$_{\odot}$ Mpc$^{-3}$. The error results from propagating errors in both the scaling factor and in $M_{\rm HI}/L$. To convert the local density to a cosmic HI density we divide by the $z=0$ critical density $\rho_{\rm c,0} = 3H_{0}^{2}/8\pi G$ 
and find:
\begin{eqnarray}
     \Omega_{\rm HI} = \frac{\rho_{\rm HI}}{\rho_{\rm c,0}} = (4.02 \pm 0.26)\times 10^{-4}h_{70}^{-1}.
     \label{omega_HI}
\end{eqnarray}

Table~\ref{omegahi_methods} summarizes our measurement of $\Omega_{\rm HI}$ with the two correction factors consecutively applied. For the smaller volume-limited sub-sample, we find $\rm C1^{v}=1.15$, and
\begin{eqnarray}
     \Omega_{\rm HI}^{\rm v} = (3.50 \pm 0.90)\times 10^{-4}h_{70}^{-1},
     \label{omega_HI}
0\end{eqnarray}
with $\rho_{L}(z=0.024) = 1.12 \times10^8 h_{70}$ L$_{\odot}$ Mpc$^{-3}$ (given by Equation~\ref{rho_l_evolution}). The result is consistent with the magnitude-limited sample, but with larger measurement error due to the smaller sample.

\begin{table}
	\centering
	\caption{The measurement of $\Omega_{\rm HI}$ with different methods. The $K_{c}$ refers to the correction factor for confusion.}
	\label{omegahi_methods}
	\begin{tabular}{ccc} 
		\hline
		Method & Formula & $\Omega_{\rm HI}(10^{-4}h_{70}^{-1})$\\
		\hline
		Measured & $\langle\frac{M_{\rm HI}}{L}\rangle\times\rho_{L}$ & 2.96 $\pm$ 0.19\\
        Luminosity bias corrected & $C1\times\langle\frac{M_{\rm HI}}{L}\rangle\times\rho_{L}$ & 4.09 $\pm$ 0.27\\
        Confusion corrected & $K_{c}^{-1}\times C1\times\langle\frac{M_{\rm HI}}{L}\rangle\times\rho_{L}$ & 4.02 $\pm$ 0.26\\
		\hline
	\end{tabular}
\end{table}

We also compute $\Omega_{\rm HI}$ in different redshift bins, with the evolved $r$-band luminosity function. Using the Galaxy and Mass Assembly (GAMA) II survey,  \citet{2015MNRAS.451.1540L} found the sample to be well-fit with luminosity ($Q$) and density ($P$) evolution parameters introduced by \citet{1999ApJ...518..533L}. The luminosity density $\rho_{L}$ can be parametrized as:
\begin{eqnarray}
     \rho_{L}(z) = \rho_{L}(z_{0})10^{0.4(P + Q)(z - z_0)},
     \label{rho_l_evolution}
\end{eqnarray}
with the Schechter luminosity function parameters in terms of magnitudes evolving as:
\begin{eqnarray}
     \alpha(z) = \alpha(z_{0}),
     \label{lf_evolution_alpha}
\end{eqnarray}
\begin{eqnarray}
     M^{\star}(z) = M^{\star}(z_{0}) - Q (z - z_{0}),
     \label{lf_evolution_M}
\end{eqnarray}
\begin{eqnarray}
     \varphi^{\star}(z) = \varphi^{\star}(0)10^{0.4P z},
     \label{lf_evolution_phi}
\end{eqnarray}
where $P = 1.0$ and $Q = 1.03$ in the $r$-band. We use the results from \citet{2003ApJ...592..819B} as the initial value for the Schechter parameters at $z_{0}$ = 0.1. The results in Table~\ref{omegaHI_redshift_table} show no measurable evolution in $\Omega_{\rm HI}$ from $z=0.038$ to $z = 0.093$.

\begin{table}
	\centering
	\caption{The cosmic HI density $\Omega_{\rm HI}$ in different redshift bins. The confusion correction has been applied.}
	\label{omegaHI_redshift_table}
	\begin{tabular}{cccccc} 
		\hline
		$\langle z\rangle$  & $\rm N_{g}$ & <\rm $M_{\rm HI}$> & <$\rm M_{\rm HI}/L$> & C1 & $ \Omega_{\rm HI}$\\
        & & ($10^{9}h_{70}^{-2}$M$_{\odot}$) & (M$_{\odot}/$L$_{\odot}$) & & ($10^{-4}h_{70}^{-1}$)\\
		\hline
		0.038 $\pm$ 0.009 & 634 & 1.73 $\pm$ 0.17 & 0.38 $\pm$ 0.06 & 1.23 & 3.92 $\pm$ 0.63\\
		0.067 $\pm$ 0.007  & 637 & 2.83 $\pm$ 0.34 & 0.21 $\pm$ 0.03 & 2.13 & 3.97 $\pm$ 0.61\\
		0.093 $\pm$ 0.007  & 621 & 3.94 $\pm$ 0.63 & 0.22 $\pm$ 0.02 & 1.92 & 3.99 $\pm$ 0.36\\
        \hline
	\end{tabular}
\end{table}

\subsection{$\Omega_{\rm HI}$ in Luminosity Bins}
\begin{figure} 
    \centering
    \includegraphics[width=8cm]{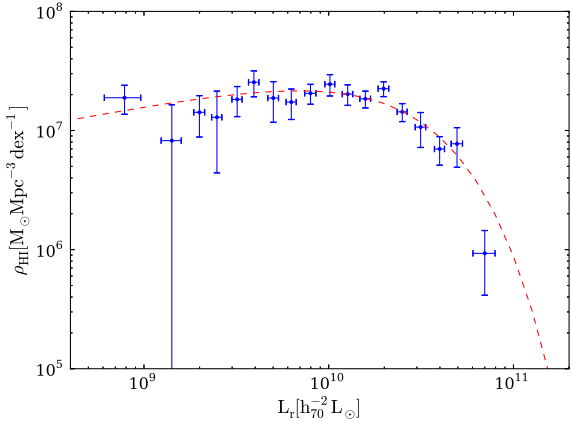}
    \caption{Values for the HI density using the measured mass-to-light ratios from the stacks as well as the luminosity density from SDSS: $\rho_{\rm HI} = \langle M_{HI}/L\rangle\times L\times\phi(L)$ The red-dashed line indicates the estimated points using fitted relation between $\langle M_{HI}/L\rangle$ and L.}
    \label{HI_density_lumibins}
\end{figure}

We also measure $\Omega_{\rm HI}$ more directly in $r$-band luminosity bins using the relation:
\begin{eqnarray}
     \rho_{\rm HI} &=&\int M_{\rm HI}(L)\phi_{L}(L)dL\\
                   &\approx&\Sigma_{i}\langle M/L\rangle_{i}L_{i}\phi_{L}(L_{i})\Delta L_{i},
     \label{density_HI_lumibins}
\end{eqnarray}
where $i$ refers to the $i$-th luminosity bin and $\phi(L)$ is the luminosity function. The $\langle M_{HI}/L\rangle_{i}$ can be obtained from Figure~\ref{stacking_lumibins}. The resultant HI density in $r$-band luminosity bins is shown in Figure~\ref{HI_density_lumibins}. Using the fits to the data and summing the density in $r$-band luminosity bins from zero to infinity, we find:
\begin{eqnarray}
     \Omega_{\rm HI} = (4.01 \pm 0.30)\times 10^{-4}h_{70}^{-1}.
     \label{omega_HI_lumibins}
\end{eqnarray}
 This is very close to the $\Omega_{\rm HI}$ derived from the stacking using the previous $\langle M/L\rangle$ bias correction. Integrating the fit in Figure~\ref{stacking_lumibins} only in the region which has data, we have $\Omega_{\rm HI} = (2.67 \pm 0.21)\times 10^{-4}h_{70}^{-1}$. If we directly sum up the data points from the stacked luminosity bins, rather than the fits, we find a value of $\Omega_{\rm HI} = (2.50 \pm 0.76)\times 10^{-4}h_{70}^{-1}$. This is lower due to the HI associated with lower and higher luminosity bins than those observed.
 
\subsection{Comparison with previous work}
\begin{figure*}
    \centering
    \includegraphics[width=16cm]{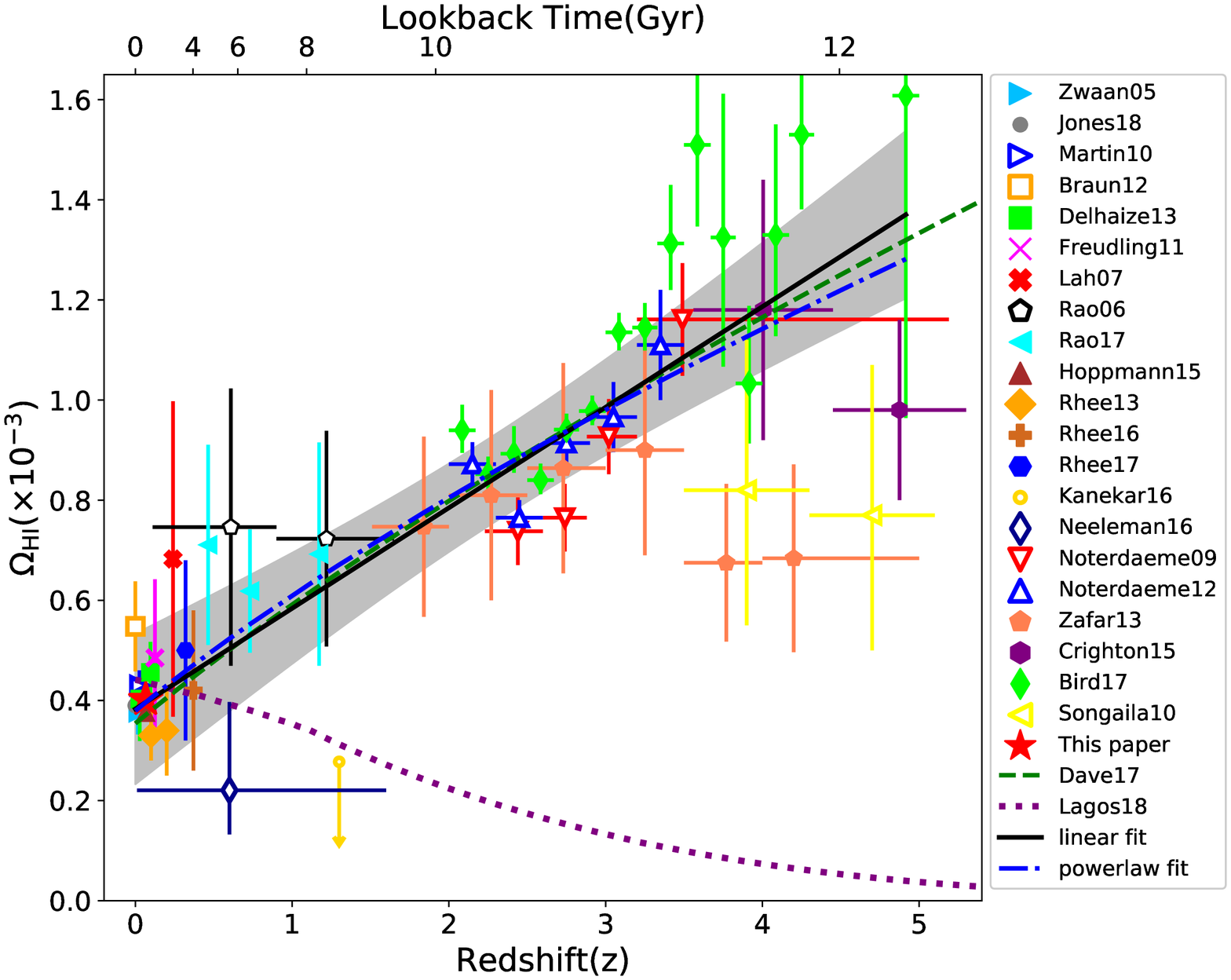}
    \caption{Cosmic HI density $\Omega_{\rm HI}$ measurements plotted as a function of redshift from different sources: HIPASS 21-cm emission measurements \citep{2005MNRAS.359L..30Z}; $\alpha40$ ALFALFA 21-cm emission measurements \citep{2010ApJ...723.1359M}; $\alpha100$ ALFALFA 21-cm emission measurements \citep{2018MNRAS.tmp..502J}; HI stacking with Parkes \citep{2013MNRAS.433.1398D}, Arecibo Ultra Deep Survey (AUDS) \citep{2011ApJ...727...40F,2015MNRAS.452.3726H}; HI stacking with WSRT \citep{2013MNRAS.435.2693R}; GMRT 21-cm emission stacking \citep{2007MNRAS.376.1357L,2016ApJ...818L..28K,2016MNRAS.460.2675R,2018MNRAS.473.1879R}; damped Lyman-$\alpha$ measurements from the HST and the SDSS \citep{2006ApJ...636..610R,2017MNRAS.471.3428R,2016ApJ...818..113N,2009A&A...505.1087N,2012A&A...547L...1N,2017MNRAS.466.2111B}; self-opaque effect corrected measurement of DLAs with GBT ans WSRT \citep{2012ApJ...749...87B}; ESO UVES measurements of DLAs \citep{2013A&A...556A.141Z}; Gemini GMOS measurements of DLAs \citep{2015MNRAS.452..217C}; measurements of DLAs with GALEX and Keck \citep{2010ApJ...721.1448S}; the MUFASA cosmological hydrodynamical simulation \citep{2017MNRAS.467..115D}; the {\sc Shark} semi-analytic model of galaxy formation  \citep{2018MNRAS.481.3573L}. Our results is shown as the red star. All of the results have been converted to a flat cosmology with H$_{\circ} = 70 $ km s$^{-1}$ Mpc$^{-1}$ and $\Omega_{\rm m,0} = 0.3$, and represent the mass density from HI gas alone, without any contribution from Helium or molecules. Missing HI from column densities below the DLA threshold is also corrected. The linear weighted fit of all $\Omega_{\rm HI}$ measurements and its $95\%$ confidence interval is shown as a black line with grey area. The blue dash-dot line shows the powerlaw fit of all measurements.}
    \label{omega_HI_plot}
\end{figure*}

\begin{figure}
    \centering
    \includegraphics[width=9cm]{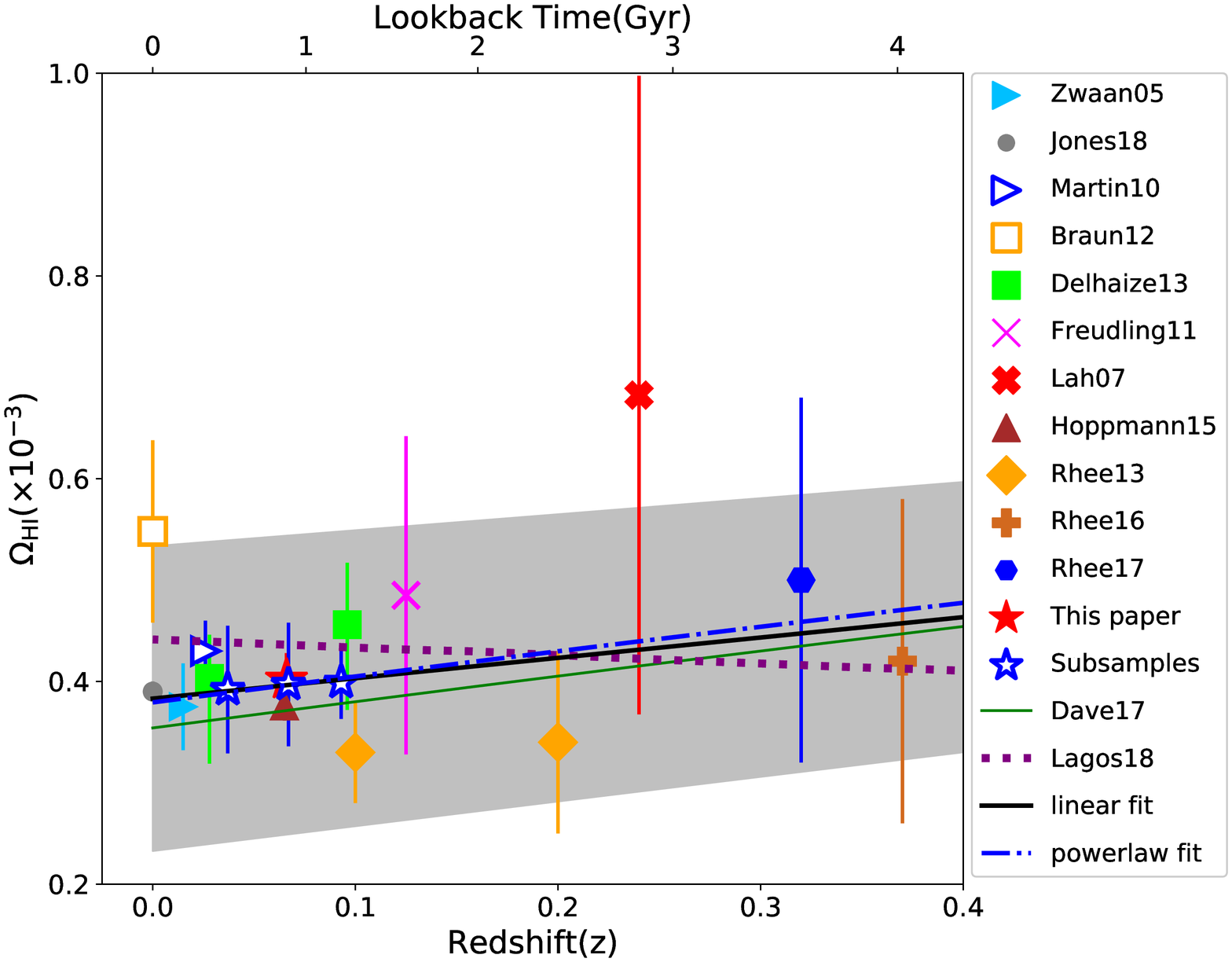}
    \caption{A zoomed-in plot showing measurements of cosmic HI density $\Omega_{\rm HI}$ from the direct HI emission and HI stacking measurements at $z<0.4$. The magenta stars represent our results from sub-samples at different redshifts. There is no discernible evolution in $\Omega_{\rm HI}$ over the last $\sim 4$ Gyr.}
    \label{omega_HI_lookback_plot}
\end{figure}

We show our results for $\Omega_{\rm HI}$ compared with other measurements at various redshifts in Figures~\ref{omega_HI_plot} and \ref{omega_HI_lookback_plot}. Each has been converted to a flat cosmology with H$_{\circ} = 70$ km s$^{-1}$ Mpc$^{-1}$ and $\rm \Omega_{m,0}$ = 0.3. $\Omega_{\rm gas}$ measurements using DLAs sometimes taken into account neutral Helium and contributions from Lyman-$\alpha$ absorbers with column densities $\log N(\rm HI) < 20.3$. We convert $\Omega_{\rm gas}$ from DLAs to $\Omega_{\rm HI}$ using $\Omega_{\rm HI} = \delta_{\rm HI}\Omega_{\rm gas}/\mu$, where $\mu=1.3$ accounts for the mass of Helium and $\rm\delta_{\rm HI} = 1.2$ estimates the contribution from systems below the DLA column density threshold.

As seen in Figures~\ref{omega_HI_plot} and \ref{omega_HI_lookback_plot}, all measurements at lower redshift ($z<0.5$) are in good agreement. But at the intermediate redshifts, measurements have large uncertainty. Our $\Omega_{\rm HI}$ measurement, marked as a red star, agrees with the measurements made at zero redshift but has a small error bar, large signal-to-noise ratio, and low systematics. It shows the usefulness of the stacking technique applied to interferometers to bridge the redshift gap between measurements using Damped Ly-$\alpha$ systems and estimates using direct 21-cm detections. 

The value we measure for $\Omega_{\rm HI}$ in sub-samples at different redshifts shows no evolution, within the errors of the measurements. In combination with other results, it again suggests almost no HI gas evolution from $z \approx 0.4$ to the present, a time span of over 4 Gyr. However, combining all measurements, there remains a clear increase of $\Omega_{\rm HI}$ at higher redshift. We should note that the `blind' HI 21cm surveys are measuring the `true' $\Omega_{\rm HI}$ with the only
assumption being that the HI 21cm emission is optically thin. On the other hand, HI stacking studies require galaxy redshifts, and are hence measuring $\Omega_{\rm HI}$ associated with galaxies detected in optical spectroscopic surveys. So high sample completeness is also required. SDSS appears to satisfy this criterion, but the under-representation of low-surface-brightness galaxies (0.1\%) and close pairs $<55\arcsec$ (6\%) may slightly skew the results, but this is not expected to be significant. $\Omega_{\rm HI}$ values from DLAs are similar to those from blind surveys, in that association of the gas with a galaxy is not a pre-requisite. However, there are a number of other biases such as dust obscuration, covering factor and lensing which may contribute uncertainty \citep{2001A&A...379..393E,2006ApJ...646..730J}. 

Many simulations have trouble reproducing the observed trend with redshift due the difficulty of resolving the various relevant gas phases (i.e. ionised, atomic and molecular gas, inside and outside galaxies). Recently,  \citet{2017MNRAS.467..115D}, using a mid-size cosmological hydrodynamical simulation, MUFASA, found $\Omega_{\rm HI} = 10^{-3.45}(1+z)^{0.74}$, which is close to the best-fit we find for the observations (Figure~\ref{omega_HI_plot}). Interestingly, previous hydrodynamical simulations have suggested that most of the HI in the Universe at $z\gtrsim 1.5-2$ is in the circumgalactic medium rather than the interstellar medium of galaxies \citep{VanDeVoort12}. Using the {\sc Shark} cosmological semi-analytic model of galaxy formation, \citet{2018MNRAS.481.3573L} were able to predict the amount of atomic hydrogen contributed by the interstellar medium of galaxies to $\Omega_{\rm HI}$, across time (see Figure~\ref{omega_HI_plot}). The contribution from the interstellar medium of galaxies decreases with increasing redshift, in a trend that is the opposite to the overall increase deduced from observations.

The large impact parameters (42 kpc for ALMA J081740.86+135138.2, 18 kpc for ALMA J120110.26+211756.2 and 30 kpc for ALMAJ123055.50-113906.4) measured for the host galaxies of high-$z$ damped Lyman-alpha systems provides some support for this scenario \citep{2017Sci...355.1285N,2018ApJ...856L..12N}. 

It also suggests that spectral HI stacking of galaxies at redshifts beyond $z\approx 0.8$ can reveal differences between the HI content of the Universe that is accounted for in galaxies and that measured through absorption lines. Future stacking experiments at higher redshifts will therefore provide unique and stringent constraints for models of galaxy formation. 

We also fit the relationship between $\Omega_{\rm HI}$ and redshift, assuming  a power law relation, and find $\Omega_{\rm HI} = 10^{-3.42}(1+z)^{0.68}$. A simpler linear fit to all $\Omega_{\rm HI}$ measurements, weighting all measurements according to their error, gives $\Omega_{\rm HI}(z) = 0.000384 + 0.0002z$. The fit is shown in Figures~\ref{omega_HI_plot}. Most of the measurements are reasonably consistent with the fit, although the HI 21cm stacking result of \citet{2016ApJ...818L..28K} and the HST archival study of \citet{2016ApJ...818..113N} lie below the trend.


\section{Summary}
In this paper we use an interferometric stacking technique to study the HI content of galaxies and confirm that there is little evolution in $\rm \Omega_{HI}$ at low redshift. Compared to previous studies, we are able to provide stronger constraints.

The data set is a 351-hr WSRT HI survey covering $\sim 7$ deg$^2$ of the SDSS sky containing 1895 galaxies with SDSS redshifts in the range $0.01 < z < 0.11$. Using measurements of the mean HI mass-to-light ratio, we were able to bootstrap from the SDSS luminosity function to provide an accurate measurement of the cosmic HI gas content. 

We have shown that interferometers such as WSRT offer significant advantages over single dish stacking measurements in terms of sensitivity, field-of-view and resolution which together maximize S/N ratio and minimize cosmic variance and confusion. 

Over all galaxies in the sample, we find an average HI mass of $\langle M_{HI}\rangle = (2.29 \pm 0.13)\times 10^{9} h_{70}^{-2}$ M$_{\odot}$ and HI mass-to-light ratio $\langle M_{HI}/L\rangle = (0.31 \pm 0.02)$ M$_{\odot}/$L$_{\odot}$. For a volume-limited sub-sample, we find $\langle M_{HI}\rangle = (0.84 \pm 0.13)\times 10^{9} h_{70}^{-2}$ M$_{\odot}$ and $\langle M_{HI}/L\rangle = (0.37 \pm 0.09)$ M$_{\odot}/$L$_{\odot}$.

We derived the cosmic HI density $\Omega_{\rm HI}$ by stacking mass-to-light ratio for all galaxies. As SDSS is magnitude-limited, many optically faint but HI-rich galaxies are missing. To correct for this selection bias, we derive a weight factor which accounts for the different mass-to-light ratios of the sample compared with an unbiased selection of galaxies. We find $\rho_{\rm HI} = (5.46 \pm 0.36)\times 10^{7} h_{70}$ M$_{\odot}$ Mpc$^{-3}$ and $\Omega_{\rm HI}  = (4.02 \pm 0.26)\times 10^{-4} h_{70}^{-1}$ at the mean redshift of $\langle z\rangle = 0.066$. For a volume-limited sub-sample, we find $\Omega_{\rm HI}  = (3.50 \pm 0.90)\times 10^{-4} h_{70}^{-1}$ at the mean redshift of $\langle z\rangle = 0.024$. We also derive the HI density from luminosity stacking and the SDSS luminosity function, finding $\Omega_{\rm HI}  = 4.01\times 10^{-4} h_{70}^{-1}$.

Rather than attempting to identify, then remove potentially confused targets, which has the effect of removing massive centrals and gas-rich satellites,
we corrected for residual confusion using a simulation. We also explore the robustness of the result to the effect of WSRT sidelobes. For both effects, the corrections were found to be small.

Finally, we split our sample in three sub-samples with $\langle z\rangle$ = 0.038, 0.067 and 0.093 and find similar results. Our results agree well with previous $\Omega_{\rm HI}$ measurements from HI emission surveys, HI stacking and DLA surveys. Taken together, the results confirm that there seems to be little evolution in $\Omega_{\rm HI}$ at low redshift.

\section{Acknowledgements}
The WSRT is operated by ASTRON (Netherlands Foundation for Research in Astronomy) with support from the Netherlands Foundation for Scientific Research (NWO). This research made use of the `K-corrections calculator' service available at http://kcor.sai.msu.ru/. We acknowledge the use of Miriad software in our data analysis (http://www.atnf.csiro.au/computing/software/miriad/). This research made use of the Sloan Digital Sky Survey archive. The full acknowledgment can be found at http://www.sdss.org. Parts of this research were supported by the Australian Research Council Centre of Excellence for All Sky Astrophysics in 3 Dimensions (ASTRO 3D), through project number CE170100013.

\bibliographystyle{mnras}
\bibliography{HI_stacking}

\appendix
\section{Stacked Spectra}
\label{sec:StackedSpectra}
\begin{figure*}
    \centering
    \subfigure{
    \includegraphics[width=17cm]{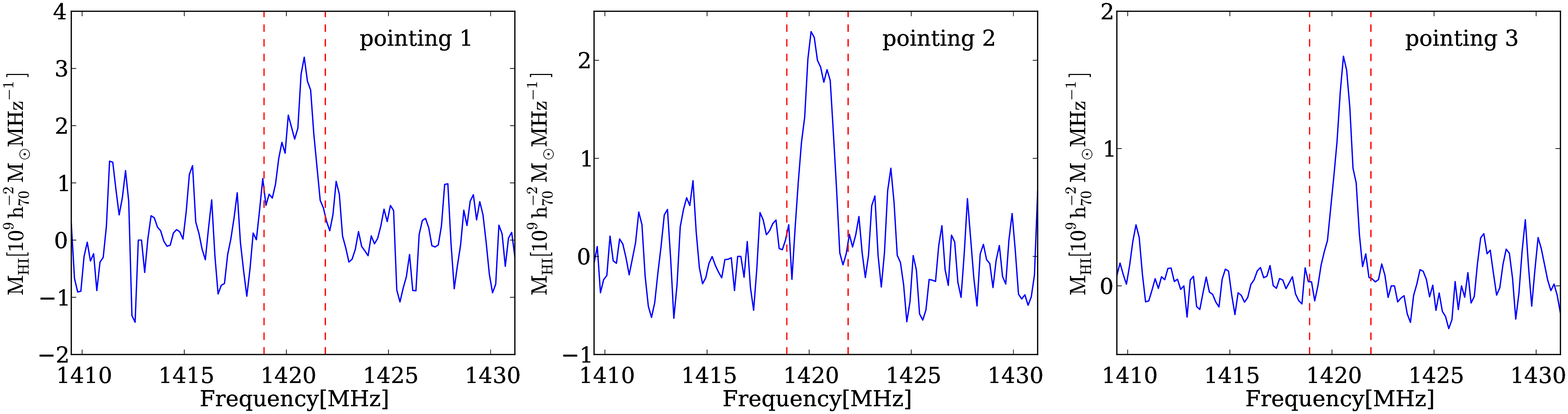}}
    \hspace{1.1ex}
    \vspace{-2.ex}
    \subfigure{
    \includegraphics[width=17cm]{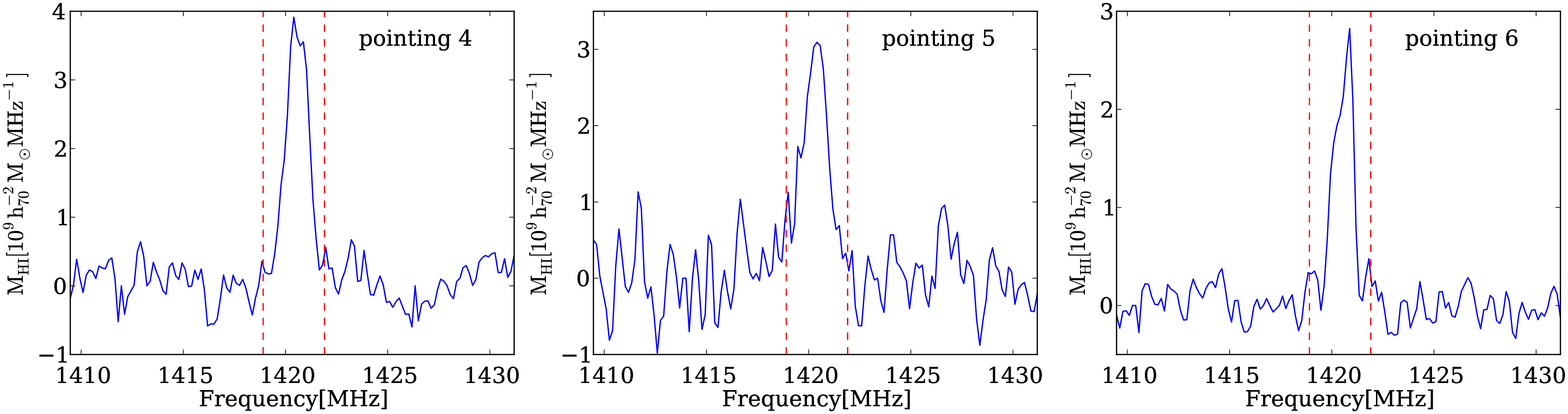}}
    \hspace{1.1ex}
    \vspace{-2.ex}
    \subfigure{
    \includegraphics[width=17cm]{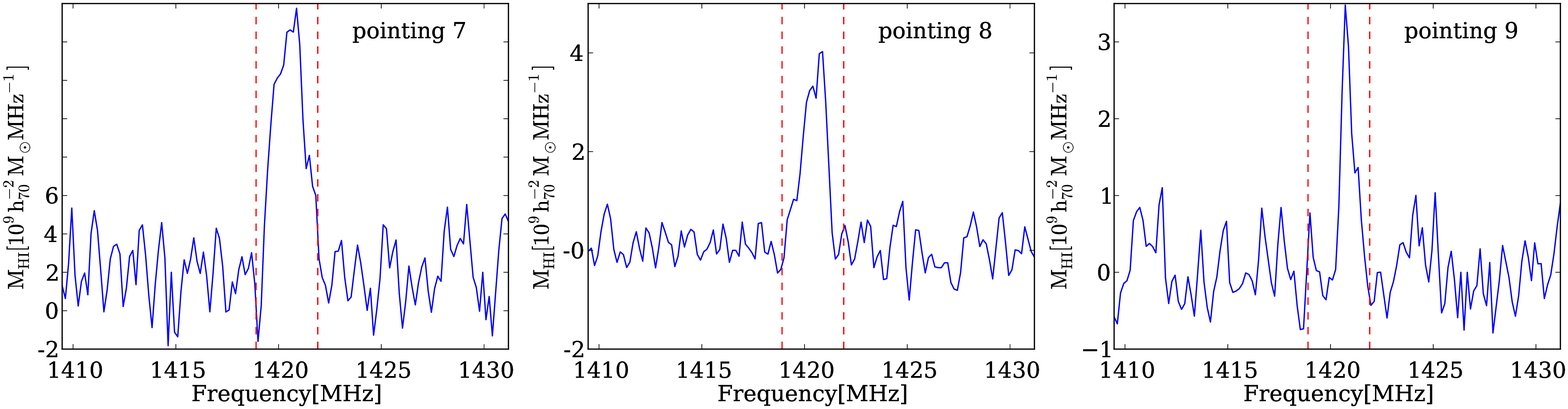}}
    \hspace{1.1ex}
    \vspace{-2.ex}
    \subfigure{
    \includegraphics[width=17cm]{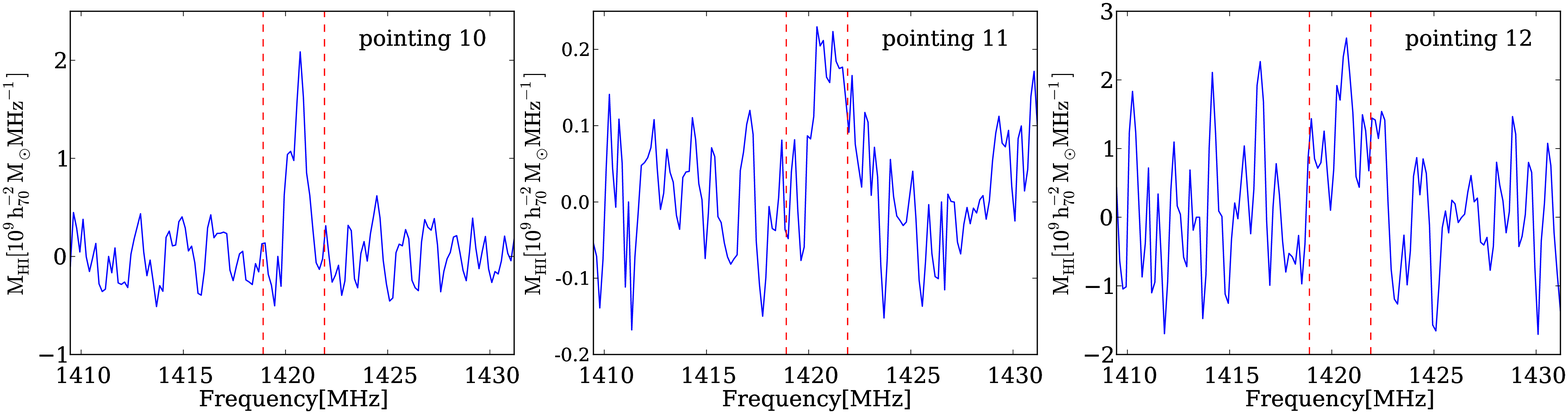}}
    \hspace{1.1ex}
    \vspace{-2.ex}
    \subfigure{
    \includegraphics[width=17cm]{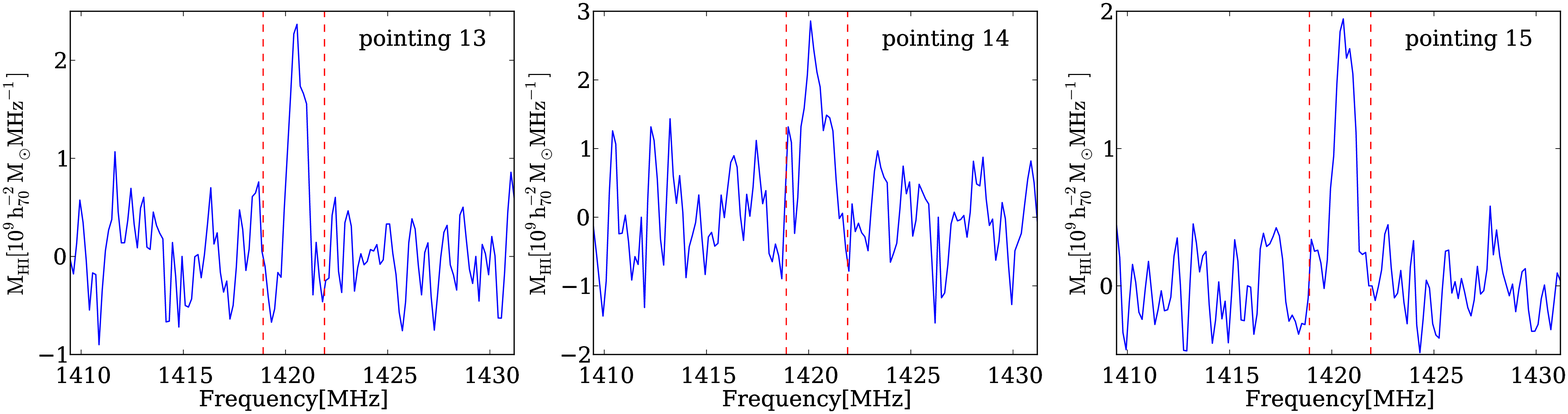}}
    \hspace{1.1ex}
    \vspace{-2.ex}
    \caption{The stacked mass spectra for pointings 1 - 15.}
    \label{stacked_spectra_1}
\end{figure*}

\begin{figure*}
    \centering
    \subfigure{
    \includegraphics[width=17cm]{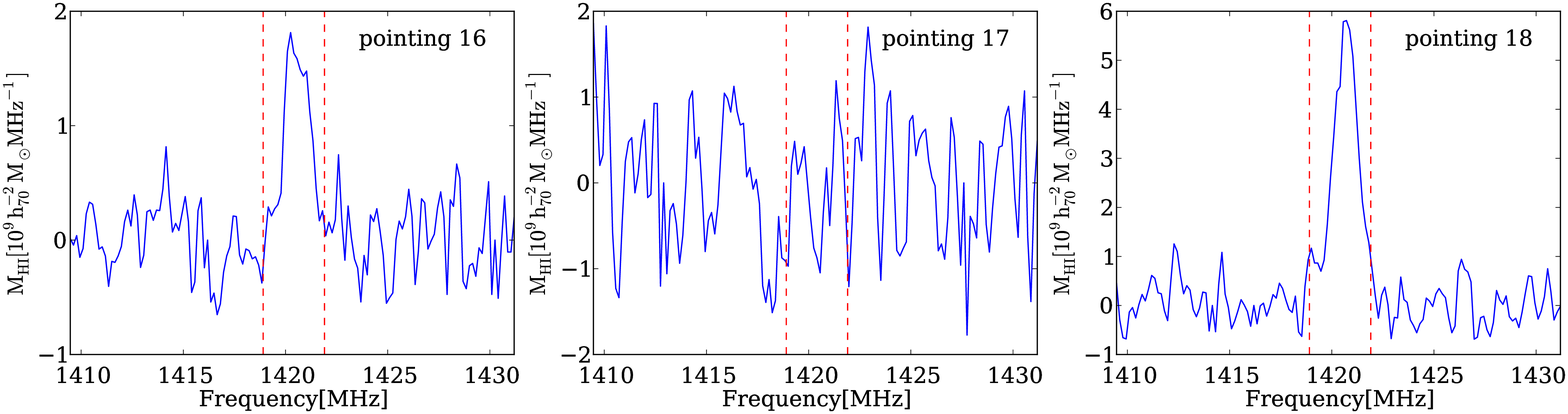}}
    \hspace{1.1ex}
    \vspace{-2.ex}
    \subfigure{
    \includegraphics[width=17cm]{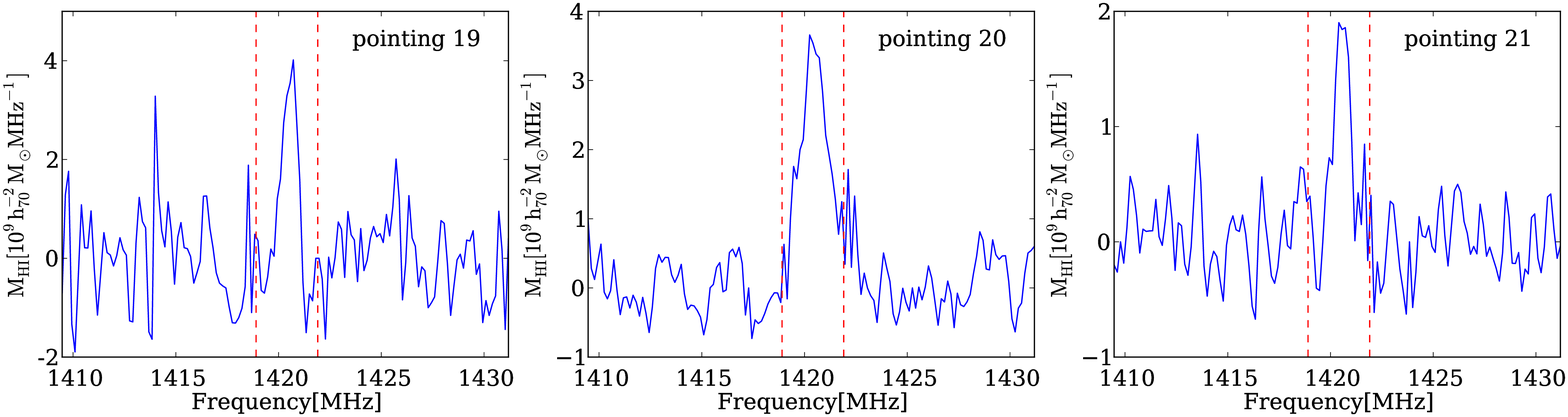}}
    \hspace{1.1ex}
    \vspace{-2.ex}
    \subfigure{
    \includegraphics[width=17cm]{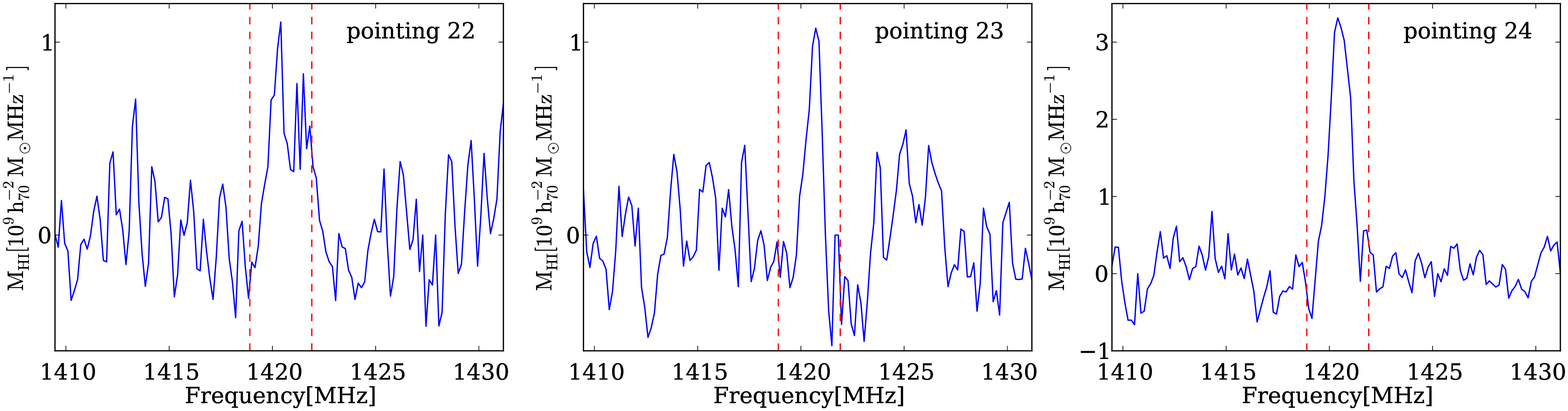}}
    \hspace{1.1ex}
    \vspace{-2.ex}
    \subfigure{
    \includegraphics[width=17cm]{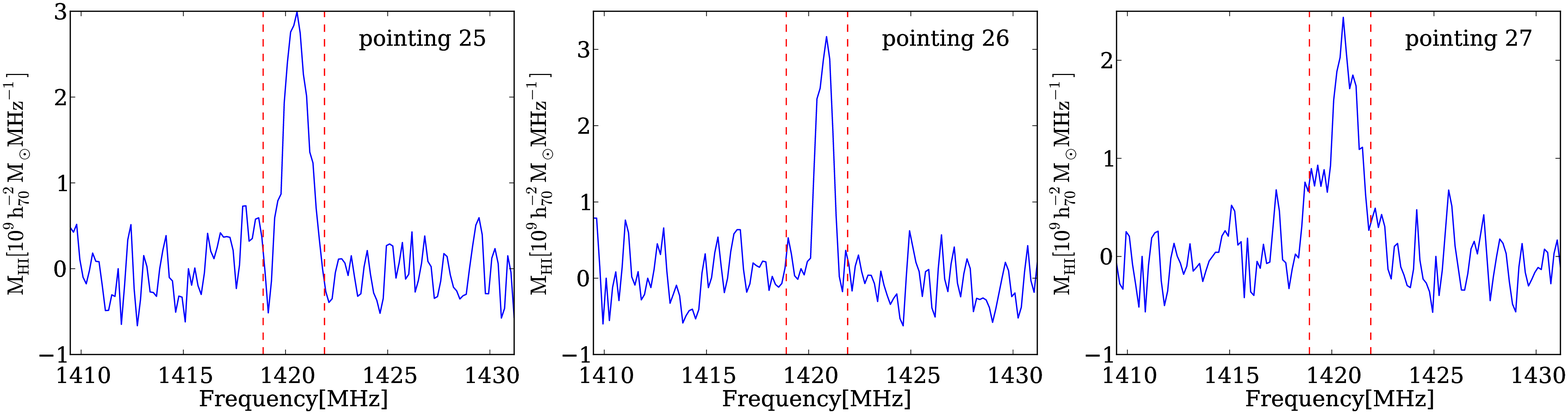}}
    \hspace{1.1ex}
    \vspace{-2.ex}
    \subfigure{
    \includegraphics[width=17cm]{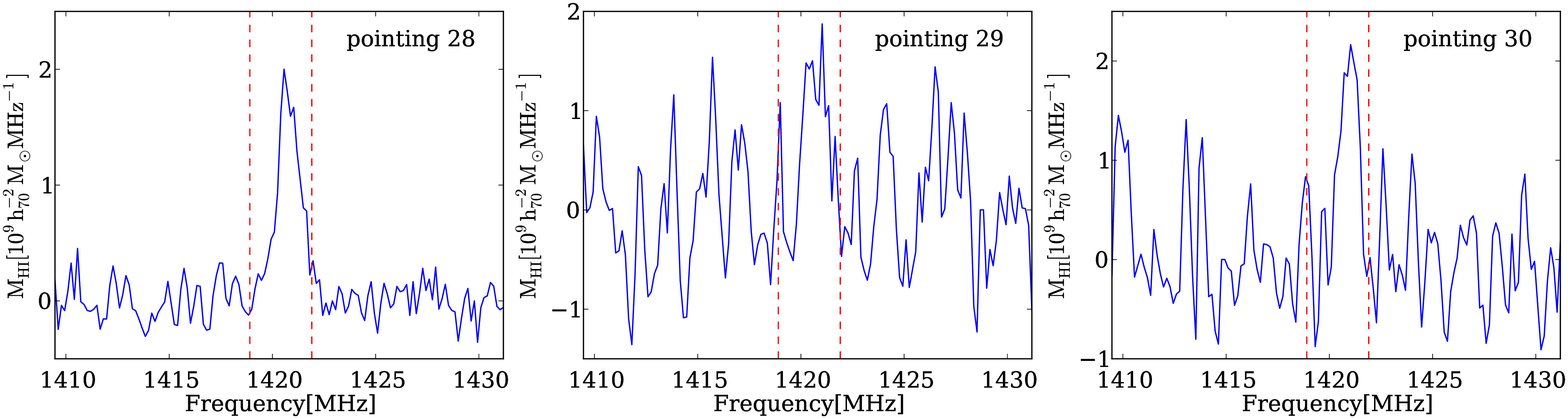}}
    \hspace{1.1ex}
    \vspace{-2.ex}   
    \caption{The stacked mass spectra for pointings 16 - 30.}
    \label{stacked_spectra_2}
\end{figure*}

\begin{figure*}
    \centering
    \subfigure{
    \includegraphics[width=17cm]{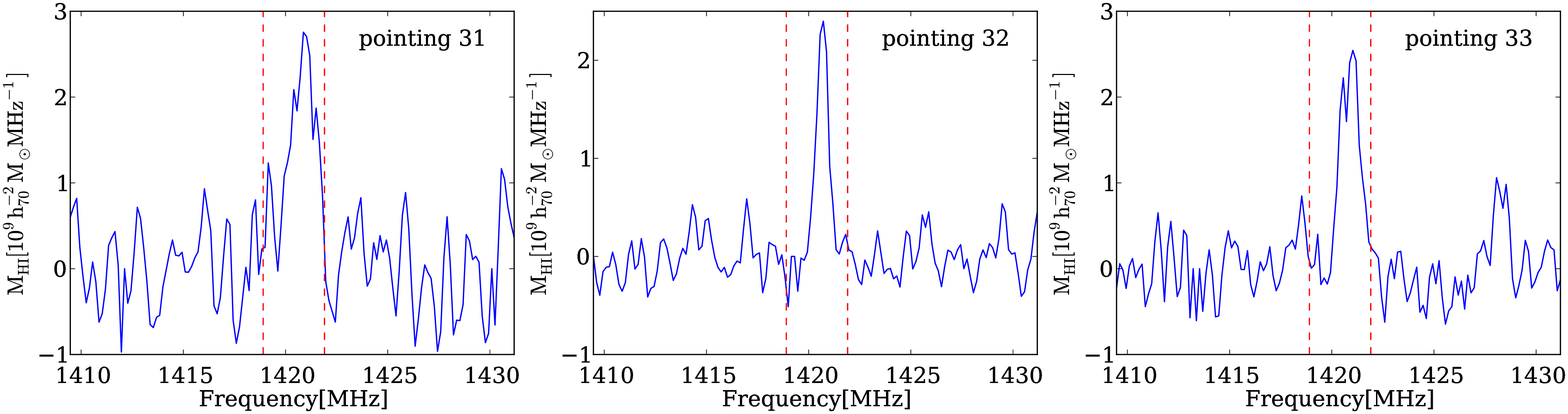}}
    \hspace{1.1ex}
    \vspace{-2.ex}
    \subfigure{
    \includegraphics[width=17cm]{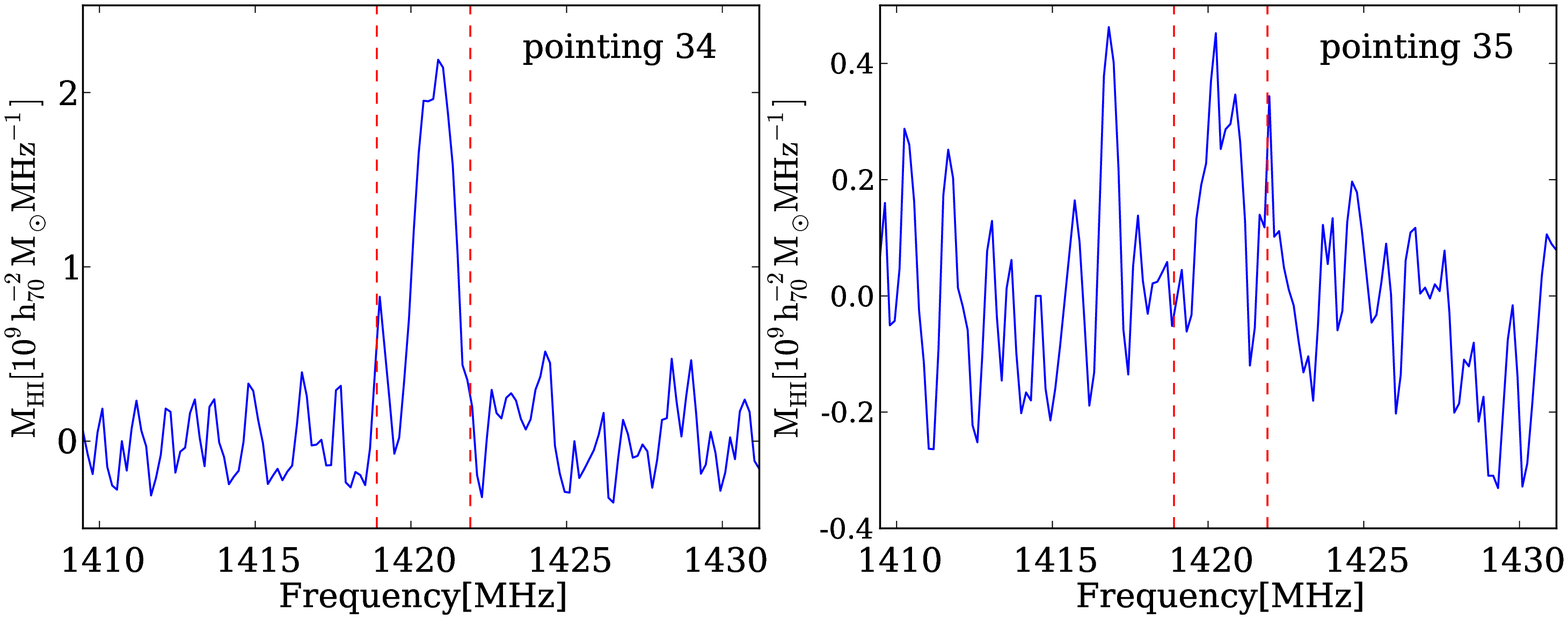}}
    \hspace{1.1ex}
    \vspace{-2.ex}
    \caption{The stacked mass spectra for pointings 31 - 35.}
    \label{stacked_spectra_3}
\end{figure*}

We show the stacked mass spectra for pointings 1 - 35 in Figure~\ref{stacked_spectra_1}, Figure~\ref{stacked_spectra_2} and Figure~\ref{stacked_spectra_3}. The red-dashed lines show the region over which we do the integration to compute the average HI mass. For the stacked mass spectra, only one stack (pointing 17) does not show a detection, three (pointings 12, 29 and 35) show unclear detections, while the remaining 30 pointings all result in clear detections.

\bsp	
\label{lastpage}
\end{document}